\documentclass[prb,aps,amsfonts,amssymb,floatfix,showpacs,superscriptaddress]{revtex4}
\usepackage{graphicx}
\usepackage{dcolumn}
\usepackage{bm}
\usepackage{color} 
\usepackage{epsfig}
\usepackage{epstopdf}
\begin{document} 
\title{Theory of Cuprate Pseudogap as Antiferromagnetic Order with Charged Domain Walls}
\author{R.S. Markiewicz and A. Bansil}
\affiliation{Physics Department, Northeastern University, Boston MA 02115, USA}
\begin{abstract}
While magnetic fields generally compete with superconductivity, a type II superconductor can persist to very high fields by confining the field in topological defects, namely vortices.  We propose that a similar physics underlies the pseudogap phase in cuprates, where the relevant topological defects, antiphase domain walls of an underlying antiferromagnetic (AFM) order, confine excess charges.  A key consequence of this scenario is that the termination of the pseudogap phase should be quantitatively described by the underlying AFM model.  We demonstrate that this picture can explain a number of key experimentally observed signatures of the pseudogap phase and how it collapses in the cuprates, while gap inhomogeneity is necessary to produce the characteristic Fermi arcs.

\end{abstract} 
\maketitle

\section{Introduction}
A major stumbling-block to understanding high-$T_c$ superconductivity is the mysterious `normal-phases' -- the pseudogap and strange-metal phases -- that give birth to superconductivity.  Why, for instance do so many strongly-correlated materials display strong intermingling/competition among multiple phases, variously labelled as ‘intertwined’\cite{IT} or ‘vestigial’\cite{ves} orders?  To what extent are these behaviors universal?

Here, we focus on the termination of the cuprate pseudogap, where recent experiments have given hints that this termination shares a number of characteristic traits among cuprates, and that these universal traits pose a strong challenge for the theory of the pseudogap.  Reference~\onlinecite{Taill1} proposes five experimental signatures of pseudogap collapse at doping $x^*$.  In this paper we set aside transport-based signatures and focus on a different set of four signatures of pseudogap collapse.  We start with two universal features from the list: (1) logarithmic specific heat divergence; (2) jump in hole doping from $x$ to $1+x$.  But a correct theory should also be able to describe non-universal features, 
and we select two that display a characteristic evolution across cuprate families: (3) the pseudogap $T^*$ collapses discontinuously, particularly for La-based cuprates; (4) doping $x^*$ of collapse varies in different cuprates, but is always close to $x_{VHS}$, where the Van Hove singularity (VHS) crosses the Fermi level.  We here demonstrate that a simple model of antiferromagnetic (AFM) short-range order collapse remarkably captures these signatures, if the role of the VHSs is appropriately taken into account.

To explain the importance of non-universality, we note that in La$_{2-x}$Sr$_x$CuO$_4$ (LSCO) the superconducting critical temperature $T_c$ is optimized at the doping $x_{opt}=x_{VHS}$, whereas in  cuprates with higher $T_c$ $x_{opt}<x_{VHS}$.  On the other hand, it is found that doping $x^*$ of pseudogap collapse is non-universal, but we shall see that it satisfies $x^*\simeq x_{VHS}$.  Indeed, cuprate-specific models are becoming more urgent, since $T_c$ itself is nonuniversal, and recent results have suggested that it scales with the fraction of holes associated with oxygen rather than copper\cite{Haase} and/or with the strength of the VHS\cite{hoVHS1}.

The early years of cuprates were dominated by the assumption that these are strongly correlated materials, so exotic theories (RVB, $t-J$ model) are needed.  More recently cuprates have been considered to have intermediate correlations ($W<U<2W$\cite{BrinkmanRice}, where $W$ is the bandwidth and $U$ is the onsite Coulomb interaction), so that more conventional many-body theory techniques may be applied.  Recently theories of short-range antiferromagnetic (AFM) order have become more popular.  There are a number of reasons for this.  First, electron-doped cuprates are widely considered to be long-range AFMs, and the picture of the resulting small Fermi surfaces (FSs)\cite{Nparm,CK} is one of the most iconic in cuprate physics.  Secondly, in the recently observed small- to large-FS crossover in hole-doped cuprates, the small-FSs can most easily be accounted for as short-range ordered versions of the same AFM order.  Thirdly, recent NMR\cite{Julien1,Julien2} and resonant inelastic x-ray scattering (RIXS)\cite{Ghir,LeTacon} experiments have shown evidence of short-range AFM (or AFM-plus-stripe) order over most of the pseudogap phase diagram.  Indeed, there is growing evidence that paramagnons associated with short-range AFM order may be the glue for high-$T_c$ superconductivity.\cite{LeTacon,Tremb1,Seamus1}
Fourthly, recent density-functional theory (DFT) calculations have shown that undoped cuprates can be quantitatively described as long-range AFMs\cite{SCAN1,SCAN2}, while optimally-doped YBCO has a number of stripe phases, which can be considered as made up of AFM domains with doped domain walls.\cite{YUBOI}  Moreover, our earlier work has shown that many-body calculations can describe many aspects of the cuprates, but only if short-range AFM order persists to high doping.\cite{AIP}  In Supplementary Material SM-A we provide a brief summary of our earlier findings, as they relate to the current work.

\section{Pseudogap collapse triggered by AFM VHS}

To explain the full phase diagram of the cuprates, we need to model the intertwined orders.  Based on the DFT results\cite{YUBOI} we consider that these are associated with doped domain walls. However, domain walls are specifically the topological defects of Ising AFMs, and in cuprates Ising order is generated by tiny anisotropies of the magnetic exchange $J$.\cite{Birg}  Thus, as the pseudogap phase boundary is approached by increasing temperature $T$ or hole doping $x$, the Ising order is expected to cross over to first $X-Y$ and then Heisenberg order, in both of which domain walls are unstable.  Hence near the phase boundary these stripe-like charge fluctuations are expected to disappear before the boundary is reached, as found in Hall effect measurements\cite{Lifshitz}.  In consequence, the pseudogap collapse with doping should be describable by a simple AFM model with charged domain walls\cite{patchmap,gapsend}.  

In this case, we find that a self-consistent calculation of the AFM collapse\cite{RMsuperVI} in LSCO follows an evolution similar to that seen experimentally.  The magnetization decreases approximately linearly with doping to a certain doping $x^*$, then drops to zero discontinuously in a first-order transition.  Moreover, the AFM stabilization energy is essentially controlled by the  Van Hove singularity (VHS) of the lower magnetic band, so that the first-order transition occurs when this VHS crosses the Fermi level and the AFM becomes unstable.  This VHS has an enhanced, logarithmic peak, consistent with the heat capacity results.\cite{Michon}

However, $x_{VHS}$ varies strongly in different families, changing the shape of the FS, and this has profound consequences on many properties.  We give two examples.  First, cuprates with small $t'/t$, close to the Hubbard model $t'=0$, are very Mott-like, with a susceptibility that always peaks at $(\pi,\pi)$ with no hint of FS nesting, while for larger $t'/t$ the susceptibility becomes more Slater-like, as the peak shifts to an incommensurate value consistent with FS nesting.  [Hopping parameters $t$, $t'$, and $t''$ are defined in Section III.]  This is not a simple crossover but a catastrophic event, with the magnetic correlation length falling to zero at the transition (an emergent spin liquid) and remaining at least an order of magnitude smaller throughout the Slater regime.\cite{MBMB}  As a second example, the VHS gets stronger with increasing $t'/t$, until it changes from logarithmic to power-law divergence, creating a high-order VHS.  There is a clear correlation between the superconducting $T_c$ and the strength of the VHS, even though $x_{opt}\ne x_{VHS}$.\cite{hoVHS1}  
Given this, one would expect to find changes in the nature of the pseudogap collapse in different cuprates.  Our calculations confirm this, producing numerous predictions for detailed testing and refining of the model.

Potential problems with a VHS scenario were quickly pointed out: (1) the nonmagnetic VHS density of states (DOS) at doping $x_{VHS}$ is logarithmic but too weak to explain the divergent heat capacity\cite{Michon} -- indeed the divergence can be completely cut off if the cuprates are sufficiently three-dimensional\cite{ZXLifshitz}; and (2) if $x^*=x_{VHS}$, then the large Fermi surface for $x>x_{VHS}$ should be electron-like with area $\sim 1-x$, rather than the observed hole-like with area $\sim 1+x$. Our calculations provide clear explanations for these potential problems.  


\section{AFM DOS peak}

\begin{figure}
\leavevmode
\rotatebox{0}{\scalebox{0.44}{\includegraphics{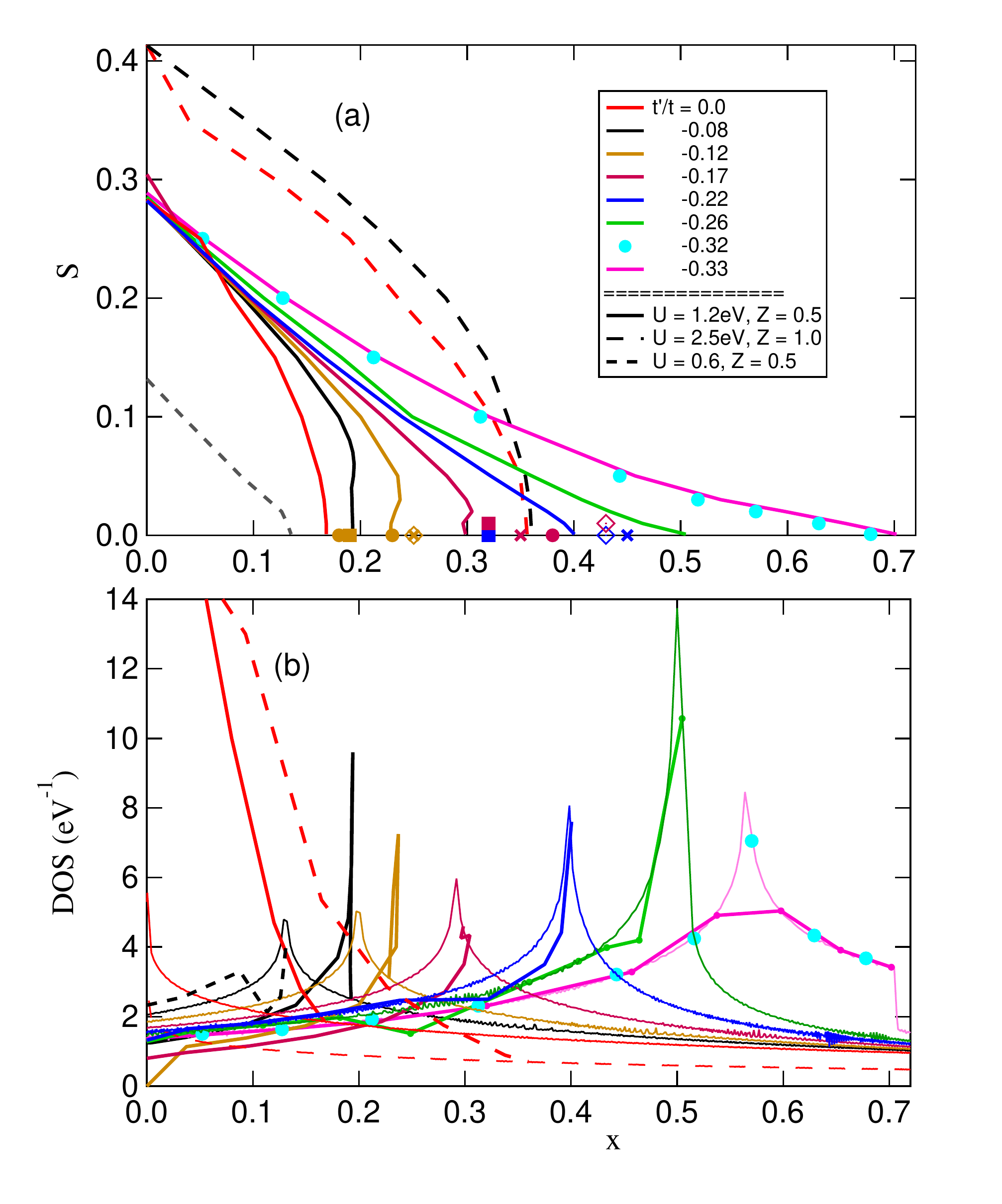}}}
\vskip0.5cm
\caption{
$(\pi,\pi)$-AFM phase in cuprates.  (a) AFM magnetization $S$ and (b) corresponding DOS as a function of doping $x$, for several values of $t'$, Hubbard $U$, and renormalization $Z$ (see legend).  In (a), the filled colored circles\cite{TaillFS} and squares\cite{OngHall,HuHall} along the horizontal axis represent experimental estimates of the pseudogap termination $x^*$ in two varieties of LSCO (brown), BSCO (dark red), and TBCO (blue), while the open diamonds\cite{OngHall,HuHall} represent $x_H$, the doping at which $n_H$ diverges.  The $X$s represent the expected $x^*$ for the model $t'$ values.
In (b) the thin lines of the same color represent the corresponding nonmagnetic phases.  In the Main Text we focus on the solid line data (plus blue dots), leaving the dashed-line data for the SM.
}
\label{fig:79}
\end{figure} 

We model the cuprates by a three-parameter tight-binding model, with the nearest $t$, the second-nearest $t'$, and the third-nearest neighbor $t''$ hopping parameters, of the form introduced by Pavarini and Andersen (PA), $t''=-0.5t'$.\cite{PavOK}  While we mainly study properties as a function of $t'$, we note the following correspondences to the known monolayer cuprate families: $t'/t \sim$ -0.13 for La$_{2-x}$Sr$_x$CuO$_4$ (LSCO), -0.20 for Bi$_2$Sr$_2$CuO$_{6+\delta}$ (BSCO), -0.245 for Tl$_2$Ba$_2$CuO$_{6+\delta}$ (TBCO), and -0.25 for HgBa$_2$CuO$_{4+\delta}$ (HBCO).  The choice of parameters is discussed in SM Section SM-II.A.

A brief discussion on the self-consistent AFM calculation is necessary, as the DFT bandwidth differs from ARPES experiments by a factor of ~2.  In a many-body perturbation study, it was found that the DFT values should be taken as the bare values, while the ARPES values correspond to quasiparticles dressed by mostly magnetic fluctuations.\cite{AIP}  Specifically, a GW calculation of the self-energy finds that a large peak in the imaginary self-energy leads to a splitting of the dispersion into a narrow coherent quasiparticle dispersion, with an incoherent residue at higher energies.  This model correctly captures many properties of cuprates, while predicting that magnetic order persists to high doping, as in the present model of the pseudogap.\cite{AIP}  Once this result is known, additional calculations can be simplified {\it a la} Landau, by ignoring the incoherent part and using the quasiparticle dispersion.  A further correction is that the Hubbard $U$ parameter
is quickly screened by doping.  The net result is that we work with DFT-based hopping parameters reduced by a renormalization factor $Z\sim 0.5$ and a Hubbard $U$ $\sim$~1.2~eV\cite{AIP}. After this we can use a  Hartree-Fock calculation, self-consistent in the doping $x$ and the dimensionless magnetization $S=<n_{\uparrow}-n_{\downarrow}>/2$, to determine the AFM phase diagrams, and the subtle role of the VHS in the lower magnetic band.  

Figure~\ref{fig:79}(a) shows the resulting phase diagram, plotting $S$ vs. hole doping $x$, while Fig.~\ref{fig:79}(b) shows a peak in the density of states (DOS) near the magnetization collapse.  Details, including alternative choices of $U$ and $Z$ (dashed lines in Fig.~\ref{fig:79}), are in SM-II.B.  In Fig.~\ref{fig:79}(a), we see that for all undoped cuprates ($x=0$) the average magnetic moment $\mu=2S\mu_B\sim 0.6\mu_B$, where $\mu_B$ is the Bohr magneton, in good agreement with experiment\cite{Tranq2,Kanun}.  In contrast, the doping dependence varies remarkably with $t'$.  Indeed, in the Mott phase the magnetic order vanishes soonest, as one is always doping away from the AFM VHS peak, causing a fast falloff of the DOS.  For all other cuprates doping the AFM phase moves the Fermi level closer to the AFM VHS peak, which helps stabilize magnetic order.

 
Our key results are that (1) the AFM gap collapses close to the point where the VHS of the lower magnetic band (LMB) of the AFM phase crosses the Fermi level.  (2) Especially for small $t'$, this VHS (thick solid lines Fig.~\ref{fig:79}(b)) has a much stronger divergence than the NM VHS (thin solid lines).  (3) In this regime the AFM gap collapse is first-order, as seen by the retrograde $S(x)$ curve, and falls at a doping beyond the NM $x_{VHS}$.  (4) The cuprates display considerable $t'$ dependence, with three crossovers.  For the Hubbard model, $t'=0$, the VHS is at the top of the LMB, so doping $p=0^+$ is always at the low-energy side of the VHS, and there is no first order transition, although the rapid decrease of DOS with doping leads to a very abrupt transition.  In contrast, for $t'/t\le-0.08$ doping is toward the VHS with discontinuous pseudogap collapse.
For $t'/t\le\sim -0.2$ the AFM transition becomes continuous and the AFM and NM VHSs converge.  In this range of $t'$ the NM VHS grows larger as it approaches a power-law divergence near $t'/t\sim -0.26$.\cite{hoVHS1}  Below $t'/t=-0.26$, a weak AFM order persists beyond $x_{VHS}$ and the DOS is virtually identical in the AFM and NM phases, as discussed in SM-II.B.  We note that the strong role of the AFM VHS in propping up the AFM order means that the phase diagram in the vicinity of this VHS can only be understood from a self-consistent calculation.  
(5) The strong $t'$ dependence of the AFM collapse is in good semi-quantitative agreement with the observed variation of the doping of pseudogap collapse $x^*$ within different cuprate families\cite{TaillFS,OngHall,HuHall} (solid colored symbols in Fig.~\ref{fig:79}(a)), and helps validate our choices of $t'$ for different cuprates.  We note however that the agreement gets worse as $x^*$ gets larger.  A plausible explanation for this is that our model is based on doping-independent $t'$ values derived from DFT dispersions of the undoped cuprates, whereas ARPES experiments find that the magnitude of $t'$ decreases with doping\cite{Yoshida,ZXLifshitz}.  Differences between experimental values of $x^*$ for Bi-cuprates may be related to the difficulty in determining the experimental doping\cite{HuHall}.  

\section{Effect of gap patches on experiment}
\begin{figure}
\leavevmode
\rotatebox{0}{\scalebox{0.44}{\includegraphics{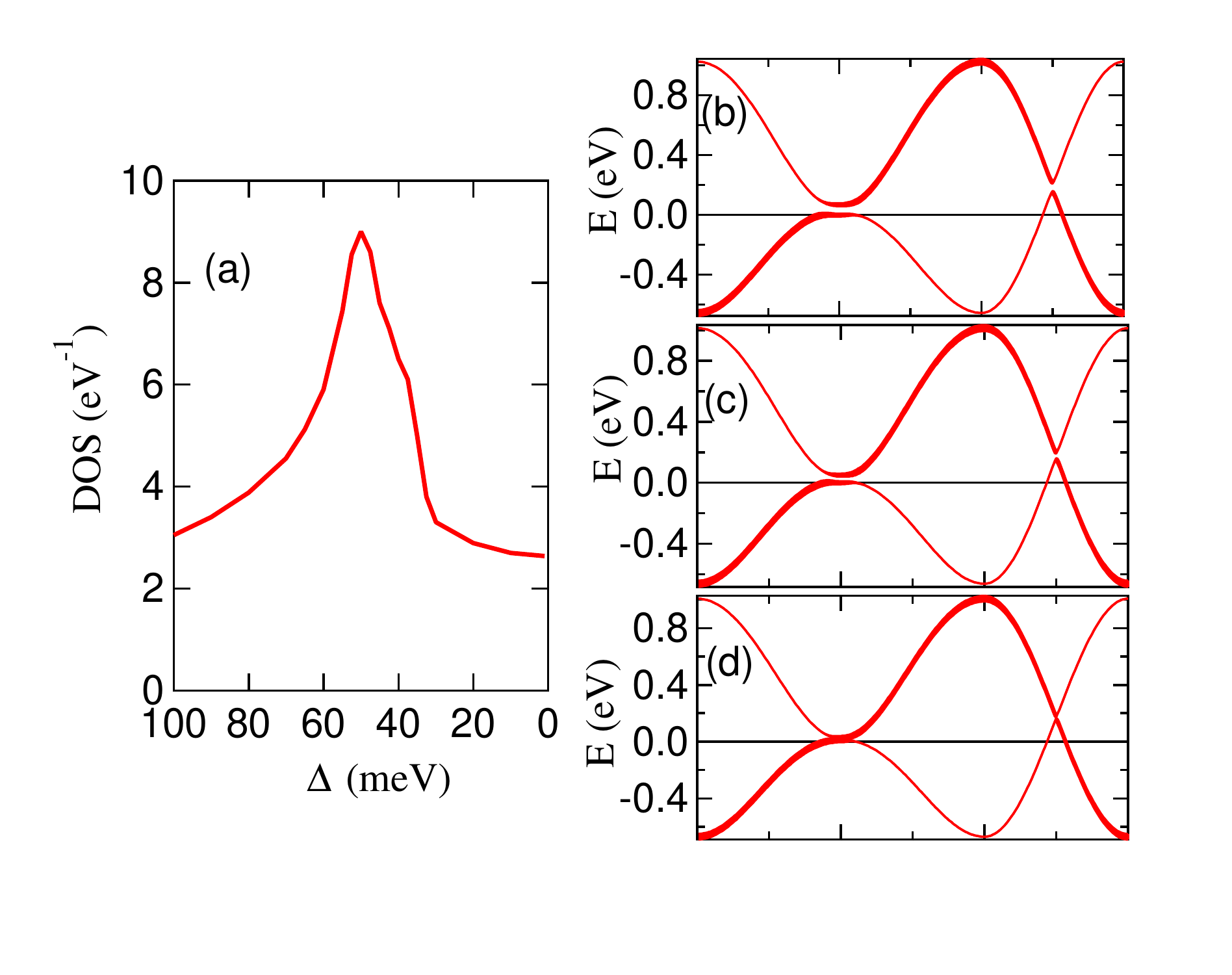}}}
\vskip0.5cm
\caption{
(a) AFM VHS, to be compared with experimental heat capacity peak, assuming that doping changes the average gap value in the underlying gap map.  (b-d) Evolution of dispersion over a similar doping range, with AFM gap = 60 (b), 40 (c) and 20 meV (c).  The thickness of the dispersion indicates the expected spectral weight of the dispersion peak, with the thinnest lines corresponding to zero intensity.
}
\label{fig:2b}
\end{figure}

These results, particularly 1-3, are in good agreement with heat capacity measurements of pseudogap collapse.  However, the careful reader might have noticed that the simultaneous presence of strong DOS divergence and first order transition would seem to be incompatible.  Possible explanations for this are discussed in SM-B.3.  We conclude that this is most likely stabilized by a form of nanoscale phase separation (NPS) [SM-A], specifically the gap patch maps seen in scanning tunneling microscopy (STM) studies.\cite{patches,EHud1,EHud2}  Here small patches with fixed doping and gap size are pinned by spatially fluctuating dopant disorder.\cite{JennyO}  In STM images, regions of a particular gap size can be masked off and studied individually, while probes such as heat capacity see an average over the patches.  The small size of the patches can frustrate first order transitions, while the patch average gap evolves smoothly with doping.   Figure~\ref{fig:2b}(a) shows the resulting AFM VHS peak.  We note that it resembles the observed heat capacity peak\cite{Michon}, in that it diverges logarithmically, is stronger than the conventional nonmagnetic VHS, and even has a similar skewness.  The fact that the AFM order is quite 2-dimensional also favors the present model of the DOS peak, as discussed in SM-B.4.

\begin{figure}
\leavevmode
\rotatebox{0}{\scalebox{0.44}{\includegraphics{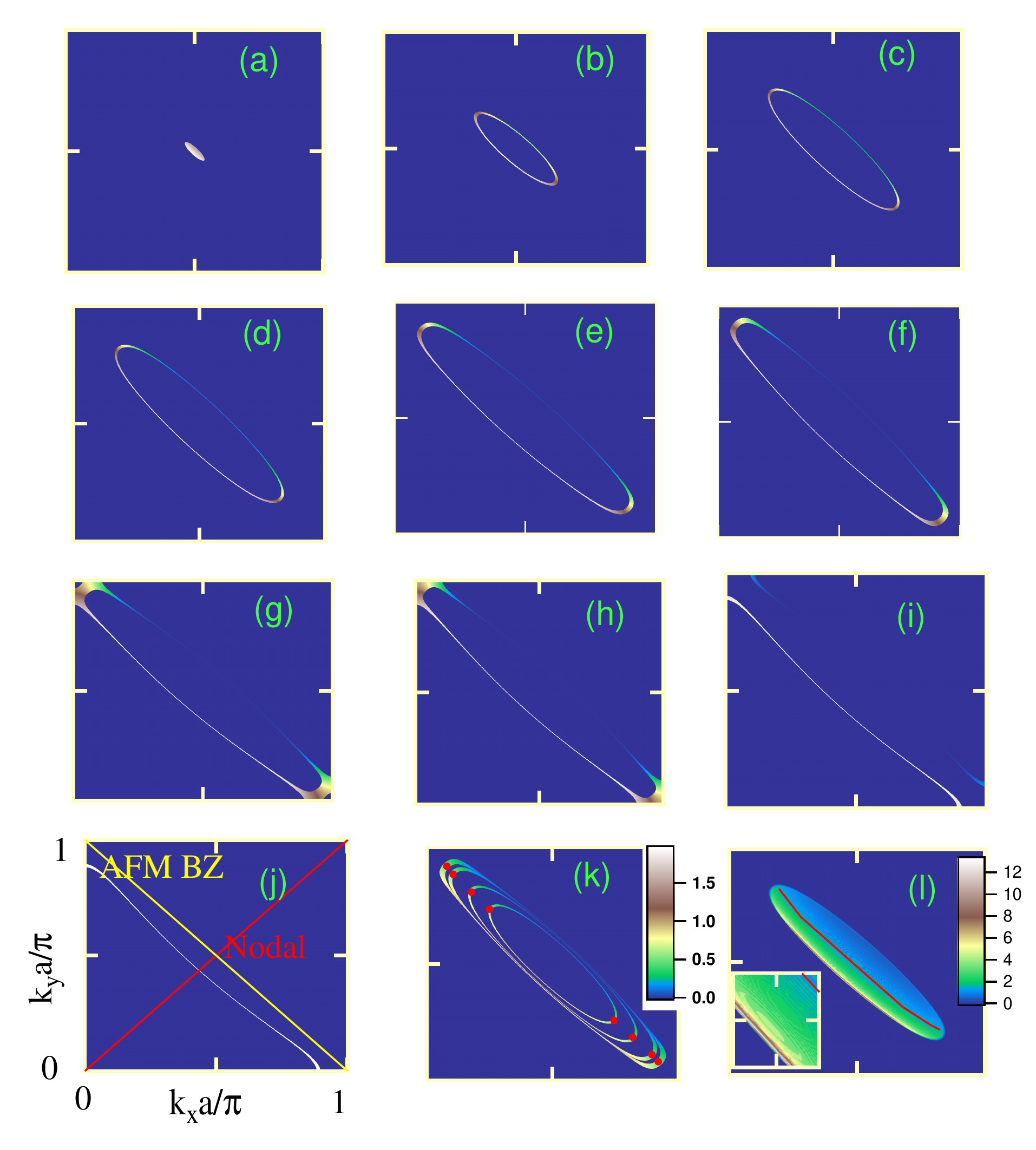}}}
\vskip0.5cm
\caption{
(a-j) Evolution of Fermi surfaces across the AFM regime, to be compared with experimental ARPES and STM data, plotted vs the average gap value.  Gap sizes are $\Delta$ = 283 (a), 250 (b), 200 (c), 150 (d), 100 (e), 80 (f), 55 (g), 40 (h), 20 (i), and 1~meV (j).  The color of the dispersion indicates the expected coherence factor on the Fermi surface, with white (blue) corresponding to a coherence factor of unity (zero).  To get a nearly complete dispersion on a finite grid of points, we count all points lying within $\pm2$~meV of the Fermi surface as lying on the surface.  This is typically better resolution than in experiment.  (k) Superposition of Fermi surfaces for $\Delta$ = 200, 150, 100, and 80~meV, to illustrate the effect of averaging over a distribution of gaps. (l) Patch map Fermi surface.} 
\label{fig:not2b}
\end{figure}

By plotting the AFM dispersion over a similar energy range, Figs.~\ref{fig:2b}(b-d), we gain insight into what ARPES experiments should see.  First, for the smallest gap (d), the gap and resulting shadow bands have nearly disappeared, leaving a dispersion nearly indistinguishable from the NM dispersion.  Secondly, even for the largest gap of 60~meV (b), any sign of the shadow bands is extremely weak, and the gap is only prominent near the VHS saddle point at $(\pi,0)$ and near $(\pi/2,\pi/2)$.  However, the latter gap is invisible to ARPES, lying well above the Fermi level, while the former feature is still above the Fermi level, making it difficult to see.  We note that a gap predominately at $(\pi,0)$ is a signature of the VHS instability.

However, we have not yet accounted for the role of the gap patches on the spectra.  Since ARPES has a similar depth penetration to STM but less spatial resolution, it should see an average over all the gap maps seen by STM.  Since the STM has such high spatial resolution, it can measure the Fermi surface within a single gap patch, to be compared with the calculations for long-range AFM order in Figs.~\ref{fig:not2b}(a-j). To aid the discussion, we note that frame (k) defines the nodal line and the AFM Brilloin zone boundary, while the nodal part of the Fermi surface lies nearest the nodal line, and the antinodal part lies furthest away, near $ka$ = $(0,\pi)$ or $(\pi,0)$.  For simplicity, we focus on the near-nodal hole Fermi surface, while neglecting small electron pockets in the antinodal region.  While we display the AFM Fermi surface over the full doping range, we caution that far from the VHS doping (e.g., frames (a-d)), there should be additional features associated with stripes or other charge order. We note a number of points of similarity with the data of Ref.~\onlinecite{SeamusD}.  First, in the nonmagnetic phase the Fermi surface can intersect the AFM zone boundary at arbitrary angles, while in the AFM phase the intersection is always at right angles, as in experiment.  Second, the long-range coherence factor is always $>1/2$ on the inside of the zone boundary and $<1/2$ outside, while the short-range order on a patch, of typical diameter 3~nm, has, for most gap sizes, large spectral weight on one side of the zone boundary and $\sim$ zero weight on the other side, as in a Fermi arc.\cite{Farc}
However, as the gap gets small, the observed Fermi surfaces extend across the zone boundary, resulting in an arc-to-Fermi surface crossover.  In the theory this crossover occurs between frames (c-f) and frames (g,h).  While at any $k$-point the spectral weight falls to 1/2 at the boundary, near the VHS the high DOS means many more points will fall near the Fermi surface.  Finally, we predict the strong changes in Fermi surface that arise after the VHS has crossed the Fermi level, frames (i,j), which have not to our knowledge been probed by STM.

With this background, we ask what ARPES should see in the patch model.  To guide the discussion, Fig.~\ref{fig:not2b}(k) illustrates what a superposition of only a few patches would look like.  We note two points.  First, along the nodal direction all curves bunch up to produce a well-defined if slightly broadened Fermi surface contour, while in the antinodal direction the curves spread out, leading to a large smearing of the Fermi surface.  This gives a very clear explanation for the experimental observation of Fermi arcs, 
where spectral features near $(\pi,0)$ are not seen, giving the impression that only an arc and not a full Fermi surface is present.\cite{Farc}  Secondly, along the AFM BZ boundary, the spectral weight remains close to 1/2.  But in ARPES the Fermi surface is defined as the point where the spectral weight falls to half of its maximum value.  Thus, the effective Fermi surface follows the nodal Fermi surface on one side, but the AFM BZ on the other side (red dots), consistent with Ref.~\onlinecite{PDJohnson}.  

Finally, in frame (l), we illustrate a patchmap Fermi surface.  We spline interpolated our self-consistent calculations to generate Fermi surfaces for 85 gap energies between 80 and 200~meV, then added them together to generate frame (l), to compare with the ARPES spectrum averaged over the gap patches.  The results are consistent with our analysis of frame (k), but still quite revealing.  If we define the Fermi surface contour by a contour of constant spectral weight, then three situations can arise.  If the constant contour is at weight = 1/2, one sees the full Fermi pocket.  Indeed the average gap is 140~meV, and the Fermi surface area is closest to that of frame (d), where $\Delta$ = 150~meV.  On the other hand, choosing a high cutoff of 12, one sees a classic Fermi arc -- see blowup in inset.  Alternatively, since there is a significant background in ARPES, it might make sense to choose a larger cutoff than 0.5.  The red curve illustrates a cutoff of 1.5, yielding an effective Fermi surface close to the experimental result\cite{PDJohnson}.  While laser ARPES studies confirm that the conventional ARPES is inhomogeniously broadened\cite{tomo}, the usual laser ARPES frequency, 6.2 eV, does not allow for study of the dispersion near the $(\pi,0)$ zone boundary.

We note that the effective Fermi surface of Ref.~\onlinecite{PDJohnson} has been interpreted as the true Fermi surface of the Yang-Rice-Zhang (YRZ) $t-J$ model.\cite{YRZ}  However, its area is only half of that of the AFM pocket.  Since the latter satisfies Luttinger's theorem, the {\it YRZ model fails to satisfy Luttinger's theorem}, making the YRZ model a highly exotic non-Fermi liquid  In contrast, we find a conventional explanation for the arcs in terms of the known gap patches.

\section{Low-field Hall effect}

Finite-$q$ phase transitions tend to be driven by FS or VHS nesting.  In the simplest situation, associated with perfect nesting, the transition gaps the full FS, so the density-wave transition is also a metal-insulator transition.  In the more common case of imperfect nesting, only part of the FS is nested, so the material remains metallic, but with a smaller FS or FSs.  The Hubbard model, which is the $t'\rightarrow 0$ limit of the cuprate family, displays perfect nesting, which is lost as $t'$ becomes finite.
In electron-doped cuprates, which are far from a VHS, there is long-range commensurate $(\pi,\pi)$ AFM order, and the small FSs have been seen experimentally\cite{Nparm,CK}.

\begin{figure}
\leavevmode
\rotatebox{0}{\scalebox{0.66}{\includegraphics{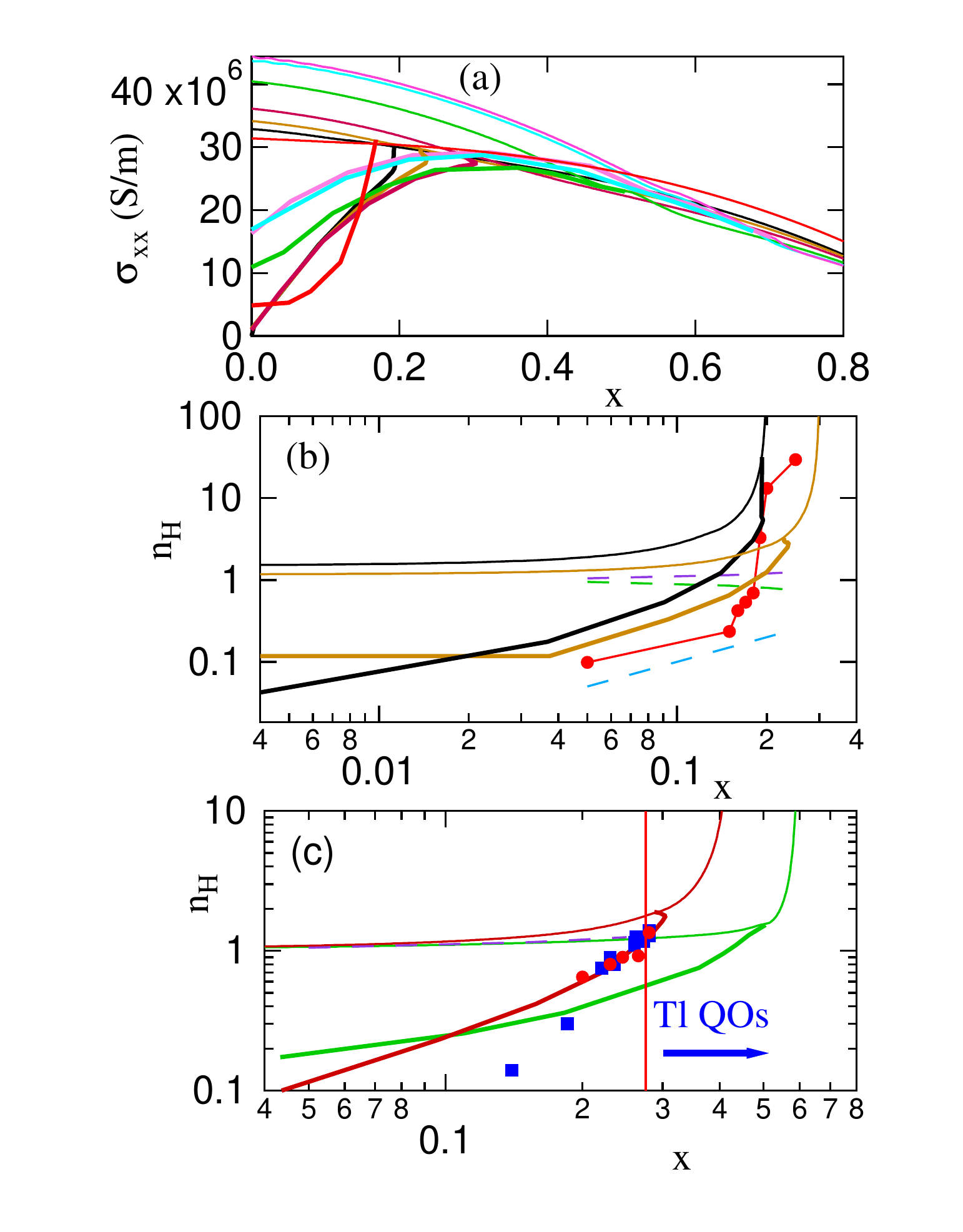}}}
\vskip0.5cm
\caption{
Low-field Hall effect compared to experiment, showing (a) magnetoconductivity $\sigma_{xx}$, for the same parameter sets as in Fig.~\ref{fig:79}. (b) Comparison of the theory for $t'=-0.08t$ (black curves) and $t'=-0.12t$ (brown curves) to experimental data on LSCO\cite{OngHall} (red dots).  Dashed lines illustrate $n_H$ scaling as $x$ (light blue), $1-x$ (green), or $1+x$ (violet).  (c) Comparison of the theory for $t'/t$ = -0.26 (green) and -0.17 (dark red curves) to experimental data on Bi- (red circles) and Tl-curates (blue squares)\cite{HuHall}.
}
\label{fig:78b}
\end{figure} 

For hole-doped cuprates, if the AFM pseudogap represents short-range AFM order, one would expect similar small FSs which should transition into a large FS at the pseudogap collapse.  If there is a sharp quasi-discontinuous transition at doping $x$ to a NM state, one expects a jump in FS area from small, $\sim x$, to large $\sim 1-x$.  However, low-field Hall effect measurements are not an ideal tool for this, since the cyclotron frequency $\omega_c$ satisfies $\omega_c\tau << 1$, so the Hall coefficient $R_H=1/n_He$ is not sensitive to the fermi surface area, but to the average curvature.

We analyze low-field transport using the Kubo formulas 
 \begin{equation}
\sigma_{ij}={-2e^2\over c\hbar^2}\int \tau\frac{\partial f}{\partial E}\frac{\partial E}{\partial k_i}\frac{\partial E}{\partial k_j}{d^2k\over (2\pi)^2},
\label{eq:1Mah} 
\end{equation}
\begin{equation}
\sigma_{ijk}={-2e^2\over c\hbar^2}\int \tau\frac{\partial f}{\partial E}\frac{\partial E}{\partial k_i}\epsilon_{krs}\frac{\partial E}{\partial k_r}\frac{\partial }{\partial k_s}\bigl(\tau\frac{\partial E}{\partial k_j}\bigr){d^2k\over (2\pi)^2},
\label{eq:2} 
\end{equation}
where $\tau$ is the scattering time, $\sigma_{ij}$ is the zero-field transport, and $\sigma_{ijk}H_k$ is the Hall conductivity with magnetic field $\vec H = H\hat z$.  [Ong has reformulated the theory to better bring out the connection to average curvature.\cite{Ong}]  The same formula is used for the NM band and the two AFM bands, except that in the latter case each point is multiplied by a coherence factor, which adds up to one when summed over the two corresponding $k$-points of the upper and lower magnetic bands (U/LMBs).  Figure~\ref{fig:78b}, along with SM Figs.~S4,5 show our results, again with the thick lines in the AFM phase and the thin lines in the NM phase.  Since we are mainly interested in the effects of band structure on transport, we simplify the calculations by working at fixed $H=$1T, $T=$100K, and assume weak impurity scattering, $\hbar/\tau=$1~meV for all curves.  

In the main text we focus on the zero field conductivity, Fig.~\ref{fig:78b}(a), and the Hall number $n_H$, Figs.~\ref{fig:78b}(b,c). Figure~S4(b) shows that for $H<$10T the cuprates are indeed in the low field limit.  The conductivity $\sigma_{xx}$ shows the same two transitions as in Fig.~\ref{fig:79}.  Thus, the finite conductivity at low doping in the Hubbard model (red curve) is associated with the high DOS VHS lying at the top of the LMB, while the next three curves with small $|t'|\le 0.26$ all start from $\sigma_{xx}\sim 0$, indicating a gap at half filling, and then follow identical growth until separation near $x\sim 0.1$.  However, for $t'=-0.26t$ the gap just closes and for smaller $t'$ the UMB and LMB overlap, leading to finite $\sigma_{xx}$ at $x=0$.  This is consistent with experiment, in that for the Bi cuprates, with $t'/t\sim -0.26$, a gap can be opened only when the Ca is replaced by a rare earth atom.\cite{BISCORE,BISCORE2}  In these calculations the resistivity $\rho_{xx}$ already displays a good qualitative agreement with experiment, SM Figs.~S5(a,c).    

The evolution of the low field Hall effect is illustrated in SM Figs.~S4(c) ($R_H$) and (d) ($n_H$),  while the FSs corresponding to $R_H=0$ are shown in SM Fig.~S5(b), illustrating the idea of zero average curvature.  Figures~\ref{fig:78b}(b,c) compare some of the $n_H$ curves, along with extra curves at different $t'$ values, with experimental data.  At a certain doping beyond $x_{VHS}$, $R_H\rightarrow 0$ and the Hall density diverges.  
In a one-band model, $R_H$ and $n_H$ should be relatively insensitive to the nature of the scattering, and the good agreement with experiment for LSCO\cite{OngHall}, Fig.~\ref{fig:78b}(b), and the monolayer Tl (blue squares) and Bi (red circles) cuprates\cite{HuHall}, Fig.~\ref{fig:78b}(c), confirms this.  

Figures~\ref{fig:78b}(b,c) constitute another key result, demonstrating nearly quantitative agreement with experiment.  Specifically, (1) for each cuprate, a value of $t'$ can be fit which accurately described the experimental data.  (2)  As expected from Fig.~\ref{fig:79}, the data for LSCO require a smaller value of $|t'|$ than for the Bi- and Tl-cuprates, leading to characteristic differences in the spectra.  Thus, in LSCO, frame (b), there is a strong increase in $n_H$ already in the AFM phase (thick curves), especially for $t'/t=-0.08$ (black curves), so that the heat capacity peak should fall at nearly the same doping as the Hall effect divergence, as observed experimentally.  On the other hand, in the Tl-cuprate, the doping dependence of $n_H$ arises mainly from the gradual closing of the AFM gap, and the $n_H$ divergence should occur at a significantly higher doping than the heat capacity peak.  Notably, in the Tl-cuprates QOs representing the large Fermi surface are found only above the red vertical line in Fig.~\ref{fig:78b}(c), quite close to the point where we predict the small Fermi surface to vanish.  Consistent with our interpretation, it was recently found that monolayer Tl-cuprates host a CDW up to a doping $x_{CDW}$=0.265\cite{Hayden}, close to our $x^*$=0.29.  This CDW could be associated with the charged domain walls in the AFM phase\cite{gapsend}.  We plot the resulting values of $x^*$ (solid squares) and $x_H$ (open diamonds) in Fig.~\ref{fig:79}(a), showing good agreement with other values of $x^*$ and with theory.  Note, however, that the values are actually derived from the best fit to the data, as $x^*$ was not reported for LSCO\cite{OngHall} while the data in Ref.~\cite{HuHall} did not go to high enough doping to determine $x_H$.  Note also that the flattening of the $n_H$ curve in LSCO, Fig.~\ref{fig:78b}(b), near the point where $R_H\rightarrow 0$, means that there is considerable uncertainty in the exact value of $x_H$.    

(3) Finally, we compare the best-fit values of $t'$ with the values  expected from the tight-binding fits to DFT spectra (SM-II.B) (denoted by $X$s in Fig.~\ref{fig:79}(a)).  For LSCO, the predicted value, -0.13, is close to the plotted data for $t'/t=-0.12$, brown curve.  On the other hand, the Bi and Tl cuprates both are fit to $t'/t=-0.17$, larger in magnitude than LSCO, but smaller than the expected $-0.20$ and $-0.245$, respectively.  As noted in connection with Fig.~\ref{fig:79}, this difference between experimental and theoretical values of $t'$ could be due to a doping dependence of $t'$\cite{Yoshida,ZXLifshitz} not accounted for in the theory.  

Thus, the present model of the pseudogap as a form of short-range AFM order, which explains the anomalous VHS peak seen at the pseudogap collapse, also captures the anomalous growth of $n_H$ with doping (albeit with some uncertainty about the optimal $t'$ for each cuprate), two key features of pseudogap collapse.  It also predicts  the nearly step-like termination of the pseudogap, particularly in LSCO.

\section{Discussion}
\subsection{Obstacles to a full theory of cuprates}

The cuprates form a large, homogeneous family of materials, but even to extract what features are universal, one needs a theory that accounts for differences between different cuprates -- especially in a theory where the VHS plays a role.  Our $t-t'-t''$ model is a compromise between highly-parametrized models that rely on very accurate FSs and a model that can be easily tuned between different cuprates with somewhat less accurate FSs.

A second key issue is the role of short-range order.  Our premise is that the theory must be developed in two stages.  First one should address the question of what is the ground state of each cuprate at each doping, and only then can one ask what are the reasons that the system does not settle down into that state, and what are the experimental consequences.  This paper is designed to mainly address the first issue, so questions of agreement with experiment must be tempered by the degree to which short-range order modifies the experimental results.  However, in Section IV we extended the analysis to account for effects of inhomogeneous broadening associated with the gap patches found in STM studies.  This allowed us to explain various pseudogap phenomena found in STM and ARPES experiments related to the observation of Fermi arcs.

We briefly discuss two further examples of heterogeneity.  First, it is widely believed that DFT calculations of nonmagnetic (NM) cuprates provide accurate descriptions of the `bare' FSs that should exist when magnetic effects are negligible.  Yet even in doped cuprates, these NM phases have energies per copper much higher than those of magnetically ordered phases.  The simple explanation of this is that cuprates maintain local moments, but these moments fluctuate so rapidly that the FS reverts to its bare value, even though the moments make a large contribution to the energy.  However, proving this assertion remains quite difficult.  There is a technique, special quasirandom structures (SQS)\cite{SpQR}, designed to deal with this problem by introducing maximally disordered moment distributions, but results are mixed.  Since the calculation is on a finite unit cell, the resulting dispersions are always discrete, with the number of bands proportional to the number of orbitals per unit cell, so improved calculations require larger unit cells.  The degree to which the dispersions tend toward a broadened version of the NM dispersion, while the energies approach a broadened average of the magnetic phases, varies widely for different materials. 
 Solving the short-range order problem would require not merely finding an accurate, converged solution of the SQS problem, but extending that solution to temporally fluctuating disorder and finite correlation-length materials.

 Secondly, on the experimental side, the various materials produced by doping La$_2$CuO$_4$ (LCO) can give us a picture of what theory is up against.  Note that we are discussing materials with the same doping, but prepared from different dopants.  Thus, adding interstitial oxygen to make La$_2$CuO$_{4+\delta}$ leads to hole-doping, but the material displays only two critical temperatures independently of $\delta$, a Neel temperature $T_N$ close to that of undoped LCO and a superconducting critical temperature $T_c=44$K, higher than that found using other dopants.\cite{Jorg}  The simple explanation for this is that, since the interstitial oxygen remains mobile below room temperature, it is pulled by the holes to produce macroscopic phase separation into separate magnetic and superconducting domains -- see Section 11.1.1 in Ref.~\onlinecite{VHSrev}.  Since other dopants substitute for La, they remain immobile, so macroscopic phase separation is frustrated, and $T_N$ and $T_c$ evolve smoothly with doping -- signatures of nanoscale phase separation\cite{VHSrev} (SM-A). Moreover, the dominant structure varies with the dopant.  For Sr doping, the material is mainly in a low-temperature orthorhombic (LTO) phase, while for Ba-doping or mixed Sr-rare earth doping, there is a broad range of low-temperature tetragonal (LTT) phase.  However, these materials also display stripe phases, where the charge stripes run along the x-axis on one layer and the y-axis on the next layer along the z-axis\cite{Tranq}.  This is interpreted as pinning of stripes to the LTT phase texture, but it only works for Sr doping if the material is locally in the LTT phase.  Such structural nanoscale fluctuations were predicted theoretically\cite{dynJT1,dynJT2} and subsequently observed experimentally\cite{Billinge,Haskel}.  

 Given all this, consider a recent experimental controversy.  A heat-capacity study of Ba-doped LCO found a logarithmically-diverging peak, which is now considered a signature of pseudogap collapse\cite{Michon}, whereas an angle-resolved photoemission (ARPES) study of Sr-doped LCO found a three-dimensional dispersion consistent with a non-magnetic material, with a non-diverging VHS DOS, which was used to match experimental heat capacity on the same material\cite{ZXLifshitz}.  Yet both materials show similar jumps in $n_H$.  How to make this data consistent?  Our interpretation is that both display pseudogap collapse, but the Ba-LCO is more ordered -- e.g., there is strong experimental evidence for nearly static stripes, which actually suppress the superconducting $T_c$, so that the diverging heat capacity is more evident.
One sees that, in the absence of a quantitative theory of short-range order, there are likely to be many seemingly incompatible experimental observations.


\subsection{Comparison with earlier work}

Some of the ideas of our model were anticipated in a short-range order model, where the holes are dressed by AFM fluctuations which produce a strongly enhanced VHS\cite{Nazar}.  
Our work also bears some resemblance to the work of Yang {\it et al.}\cite{YJarrell1,YJarrell2}.  They found that superconductivity is significantly enhanced near the cuprate quantum critical point (QCP) due to an enhanced (power-law) divergence in the particle-particle susceptibility\cite{YJarrell1}.  They found that this could be explained by a VHS at the QCP, but only if the VHS is a high-order VHS with power-law $p\sim 0.5$.\cite{YJarrell2}  However, the VHSs in NM cuprates are material dependent, varying from logarithmic to power law, but with $p\sim$ 0.25 (analytically, near the VHS\cite{Zehyer}) or 0.29 (numerically, over a broad energy range\cite{hoVHS1}).  In contrast, Michon {\it et al.}\cite{Michon} find a heat capacity consistent with an enhanced but logarithmic VHS.  Our work can reconcile these two results: we find that the AFM VHSs are close to a hoVHS with power law $p\sim$ 0.5\cite{hoVHS1}, but at a slightly different doping.  This means that at energies far from the VHS they grow with a power law $\sim 0.5$, but as they approach the VHS they break off to a slower, logarithmic growth.  

A number of cluster extensions of dynamical mean-field theory~\cite{CDMFT1,CDMFT2,CDMFT3} calculations have studied the cuprate pseudogap problem recently.  They generally find that $x^*\simeq x_{VHS}$ for small $|t'|$\cite{RMsuperVI,AIP}, but that $x^*<x_{VHS}$ for larger $|t'|$, inconsistent with experiment\cite{TaillFS} (see SM Section SM-B.4).  Indeed, they find that the pseudogap T$^*$ vs $X$ curve is virtually independent of $t'$, with $x^*\sim 0.15$, inconsistent with Figs.~\ref{fig:79},\ref{fig:78b} (both theory and experiment) -- see in particular Fig.~1(b) of Ref.~\onlinecite{CDMFT3}.

In an important paper, Kowalski {\it et al.}\cite{Tremb1} provided strong evidence that paramagnons provide the pairing glue in cuprates.  Here we note merely that their three band model, with cellular dynamical mean-field theory, is in good agreement with the three-band model with AFM order.\cite{RM70,RIXSPRL}  We note in particular that in the latter model, magnetic gaps open up in both the antibonding band at the Fermi level and in the bonding band several eV below it; the gap near the Fermi level is consistent with the AFM gap in the one-band model; and the results of this three band model have now been confirmed by first-principles calculations\cite{SCAN2}.  Notably, the band labeled Zhang-Rice singlet in Ref.\onlinecite{Tremb1} is the lower magnetic band of the antibonding band of the AFM three-band model.

\subsection{Fingerprinting cuprates}

In SM Section SM-B.1, we introduced a modified Pavarini-Andersen model of the cuprate reference family: 
\begin{equation}
r=cos(k_{nx}a)/2,
\label{eq:3} 
\end{equation}
\begin{equation}
t'=-brt,
\label{eq:4} 
\end{equation}
\begin{equation}
t''=-t'/2,
\label{eq:3} 
\end{equation}
where $k_n$ is the vector from $\Gamma$ to the Fermi surface along the nodal direction, $\Gamma\rightarrow (\pi,\pi)$, at the doping $x_{VHS}$, and $b$ is a constant in the range $\sim 2/3 - 3/4$.

We note that $r$ is a {\it fingerprint for each cuprate,} and a signature of the deviation from the electron-hole symmetric Hubbard model, and is easy to measure from either experiment or first-principles calculations.  Moreover, $b$ can be found from fitting data to the $t=t'-t''$ model, as discussed above, and comparing the $t'$ and $r$ values.  Accumulating such data could put the study of cuprates on a much more quantitative basis, and answer some fundamental questions, such as:  Are the $r$ values the same for theory and experiment? (Equivalently, is the cuprate self energy momentum dependent?)  Is $b$ universal, or does the off-CuO$_2$-plane structure modify the dispersion?  Similarly, for La- and Bi-cuprates, do rare-earth substitutions modify $r$ or $t'$?  Which one?  What physical properties scale with $t'$?  We know that $T_c$ does\cite{PavOK}, does the fraction of oxygen holes\cite{Haase}?

\subsection{Topological Defects}
Whe have focused in the present manuscript mainly on the regime of pseudogap collapse, where domain walls play a minimal role, and are leaving the study of domain walls per se to a follow-up manuscript.\cite{gapsend}  Here we would briefly like to put the issue in context, and recall earlier work on domain walls.  Topological defects are deviations from perfect lattice order, stabilized by the lattice geometry, such as vortices and domain walls.  Having reduced free energy, they can stabilize the lattice from competing phases by acting as a sink for states that would otherwise cause a large reduction of the order parameter -- e.g., excess charges confined on domain walls of an AFM, excess magnetic fields confined in vortices of a superconductor.

They came into prominence in cuprates when Tranquada found stripe phases which involved ordered arrays of alternating charge and magnetic stripes, and could be interpreted as an AFM with periodic arrays of hole-doped domain walls.  While the stripes are generally assumed to play a minor role as a secondary order in the pseudogap, largely confined to La-cuprates at low doping, $x\le 1/8$, in Ref.~\onlinecite{gapsend} we demonstrate that they have a much wider role.

\section{Conclusions}

Correlated materials combine aspects of great complexity (intertwined orders) with with other aspects hinting at great underlying simplicity -- universality, emergence of exotic phases with `colossal' properties (high-$T_c$ superconductivity, colossal magnetoresistance, heavy fermion behavior), pseudogaps, and strange metals with anomalous transport (linear-in-$T$, -$\omega$ resistivity).  For cuprates, our earlier studies (with colleagues) of MBPT extensions of DFT identified the underlying simplicity as a form of short-range AFM order which gives rise to the pseudogap \cite{AIP}, and the complexity (lack of long-range order) as due to frustration 
from lower-dimensional fluctuations and competing instabilities\cite{MBMB}.  Parallel DFT studies found the important role of topological defects of the AFM order (domain walls) in generating anomalous stripe and CDW phases\cite{YUBOI}.  Here we have shown that the same models can explain key features of the pseudogap collapse.

In this work, we focused on describing the underlying band structure effects, neglecting as far as possible the effects of heterogeneity, disorder, and scattering.  This provides a framework upon which these factors can be subsequently incorporated.  This is particularly important for the scattering, both because the effects of heterogeneity and dopant disorder will make it hard to extract the intrinsic scattering and because the electron-electron scattering is likely to be strongest near the VHS, making self-consistent calculations particularly difficult.

Given these simplifications, our model still captures many key features of the pseudogap collapse.  We model the pseudogap collapse as a collapse of the underlying AFM order.  We find striking agreement with several of the key signatures of pseudogap collapse, including the logarithmically diverging heat capacity associated with an AFM VHS, beyond which $n_H$ remains positive and increases with $x$, although it is not directly the signature of a large fermi surface.  Proximity of $x_{pg}$ to a strong VHS suggests that the linear-in-$T$ resistivity is associated with strong scattering at the VHS.  Moreover, the often observed discontinuous drop of $T^*$ to zero at $x_{pg}$ also follows naturally from our model.  A straightforward extension of the model to include gap inhomogeneity leads to a simple interpretation of Fermi arc effects.


\section*{Acknowledgements}     
This work was supported by the US Department of Energy (DOE), Office of Science, Basic Energy Sciences Grant No. DE-SC0022216 and benefited from Northeastern University’s Advanced Scientific Computation Center and the Discovery Cluster.   
We thank Adrian Feiguin and Ole Andersen for stimulating discussions.  
\section*{Author contributions}
R.S.M. and A.B. contributed to the research reported in this study and the writing of the manuscript.
\section*{Additional information}
The authors declare no competing financial interests. 
\end{document}


\title{Theory of Cuprate Pseudogap as Antiferromagnetic Order with Charged Domain Walls\\
Supplementary Material}
\author{R.S. Markiewicz and A. Bansil}
\affiliation{ Physics Department, Northeastern University, Boston MA 02115, USA}

\maketitle
\beginsupplement
\section{Review of short-range AFM theory}

Cuprates appeared at an opportune moment in physics, when many-body theories were taking advantage of DFT results to transition from theories based on the free electron model\cite{Mahan} to real-material theories based on accurate dispersions.  This was particularly important for cuprates, since the older DFT theories could not stabilize the magnetic order, but was of more general importance since doping is hard to incorporate in DFT, particularly the role of random dopant distributions.  Surprisingly, this systematic approach was largely eschewed in favor of more exotic theories, partly because such theories were thought necessary to describe Mott physics, but perhaps also because the cuprates did not seem to display any conventional transitions to long-range order.

On the other hand, cuprates are mainly two-dimensional materials, where fluctuations could profoundly modify phase transitions, as in Mermin-Wagner physics, and where a highly anomalous saddle-point VHS, which typically displays a logarithmic divergence in 2D, had already been observed.  Moreover, cuprates had only a single band at the Fermi level, eliminating many complications of many-body theory.  Reasoning that a many-body approach to cuprates would be profitible, if only to hone the theory on a real material, and to understand the competition between FS nesting and VHS nesting\cite{RiSc}, our group embarked on such a theory.

We began with a survey of the Van Hove scenario.\cite{VHSrev}  While a number of such surveys exist, ours focused on factors which could complicate the simplest picture of an ideal VHS driving a single phase transition.  Our main findings are: (1) the cuprates already display competition of several instabilities, as both AFM order and superconductivity can be enhanced by the strong DOS peak.  Subsequently, we developed an SO(8) Lie theory of competing VHS instabilities, leading to 28 possible phases, if the nesting vectors are restricted to $\Gamma$, $(\pi,0)$, $(0,\pi)$, and $(\pi,\pi)$.\cite{SO8} 

(2) While electronic phase transitions can be first order, as found in photoexcited electron-hole droplets in semiconductors\cite{Keldysh}, such phase separation is inhibited in cuprates, as the compensating negatively-charged dopant ions are generally locked to the lattice, and cannot phase separate with the holes.  This leads to a frustrated phase transition, with nanoscale phase separation (NPS)\cite{VHSnps,phassep1,phassep2}, which further complicates the search for conventional long-range order.  We find that such NPS is involved in the formation of the stripe phases found in cuprates.

(3) Not all VHS are created equal.  Clear evidence was found for stronger VHSs, some, called high-order VHSs (hoVHSs), with power law rather than logarithmic divergence.\cite{Andersen,Zehyer,RMsuperVI}  These have recently become popular, when they were found in twisted bilayer graphene and other materials\cite{Ref2a,Ref2b}, inspiring a reassessment of their role in cuprates.\cite{hoVHS1}  

This last point in particular focused on the need for numerically extending many-body theory results to realistic band structures, starting with the Lindhard electronic susceptibility\cite{Lindhard}.  This study revealed that the susceptibility contains an approximate, folded map of the cuprate FS.  The peaks in this map are typically nonanalytic at $T=0$, and give a precise meaning to the concept of FS nesting.
However, the role of additional matrix element effects beyond simple Lindhard theory is currently under debate, particularly in the case where electron-phonon interactions are involved\cite{Mazin}.  Moreover, a subsequent study found that this map is superimposed on a smooth background peaked at $(\pi,\pi)$, associated with VH nesting.\cite{MBMB}  This background has the effect of favoring nesting features near $(\pi,\pi)$, and thereby greatly extends the subtle role of the VHS on cuprate physics.  

Armed with the susceptibility results, the many-body study proceeded in two directions.  First, the susceptibility was used to calculate a GW self energy, which describes the effect of dressing holes with spin fluctuations.  This was found to split the electronic dispersion into coherent (at low energy) and incoherent (at high energy) branches.  In the incoherent branches, a residual Mott gap persists to high doping, while the coherent part acts as in-gap states, with the renormalized dispersion in good agreement with ARPES experiments.  Secondly, these renormalized bands could be used in (ideally self-consistent) random-phase approximation (RPA) calculations, as had already been applied to the magnetic order.  An early result was describing the AFM phase in electron-doped cuprates\cite{Kusko}, while more recent work focused on the more complicated hole-doped cuprates.  These results are summarized in Ref.~\onlinecite{AIP}.

The above transitions were studied using the RPA, and so are mean-field transitions to a long-range order.  The role of fluctuations has now been accounted for in a series of publications, based on a Moriya-type approach involving mode-coupling vertex corrections.\cite{RM70}  The latest paper\cite{MBMB} provides a stable implementation of the formalism involving a susceptibility DOS.  This DOS quantifies the number of modes that are competing for the phase transition, and leads to two forms of frustration.  The first is conventional Mermin-Wagner\cite{MW} fluctuations that, e.g., allow a Heisenberg AFM order only at $T=0$, while the second quantifies McMillan's concept of bosonic entropy\cite{McMill}, which can greatly reduce the ratio of $T_c/\Delta$, where $T_c$ and $\Delta$ are the transition temperature and gap for any particular ordered state.  This is McMillan's definition of a strongly-correlated material.


Applied to cuprates, this formalism gives rise to a striking non-Landau phase transition, as a function of hopping parameter $t'$.\cite{MBMB}  The folded FS map forms a diamond shape centered on ($\pi,\pi$).  Due to Pauli blocking, the susceptibility is small outside of this diamond but large inside it, leading to a susceptibility plateau.  For small $t'$ the plateau is confined close to $(\pi,\pi)$, and the tails of all the nesting features superimpose so the susceptibility always peaks at $(\pi,\pi)$, independent of $t'$ or doping.  Thus, magnetic order arises only at ($\pi,\pi$ ), and due to a weak Ising anisotropy the low-temperature topological defects are charged domain walls with purely repulsive interactions, leading to a stripe phase.  This is a clear many-body representation of a Mott phase, where FS nesting is absent.  Increasing $t'$ causes the FS to grow, the plateau to spread further away from $(\pi.\pi)$, and the overlap of the susceptibility tails to rapidly decrease.  At some point the $(\pi,\pi)$ susceptibility crosses over from a maximum to a local minimum, in the process exposing the (incommensurate) FS nesting peaks.  This leads to a more conventional Slater phase, although the reduced susceptibility intensity coupled with increased frustration associated with different $k$-points on the FS diamond lead to a reduction in correlation length by at least an order of magnitude across the transition.  However, that is not all.  When the $(\pi,\pi)$ susceptibility crosses over from a maximum to a local minimum, it passes through a state of maximal frustration, where the susceptibility is essentially flat across the full susceptibility plateau.  In this case the magnetic correlation length collapses to nearly zero, leading to an emergent spin liquid state.  A similar result was found in a 3D Hubbard model.\cite{Kohn}  In a recent reappraisal, it was noted that if the susceptibility is interpreted as the Green's function of an electronic boson (electron-hole pair), then the flat susceptibility corresponds to a bosonic hoVHS.\cite{hoVHS1}

In 2017, we started a new approach to the cuprates and other correlated materials. In collaboration with J.~Sun, now at Tulane, we showed that DFT calculations using the SCAN exchange-correlation potential can provide accurate descriptions of the magnetic state of undoped cuprates, consistent with our many-body calculations.\cite{SCAN1,SCAN2}  Moreover, in YBCO$_7$ we found over 20 low-energy phases, most based on charged domain walls (stripes)\cite{YUBOI}, that give insight into what intertwined orders might look like in a first-principles calculation.  In particular, these phases tend to bunch up at an accumulation point near the ground state, all containing large average magnetic moments.  Notably, we have now found stripe phases with a similar accumulation point in the infinite layer superconducting nickelates, even though many features of the stripes are quite different, and associated with multiband effects.\cite{nick}

\section{AFM phase transition}
\subsection{Modified Pavarini-Andersen model}

Over the past 30 years we have developed a set of accurate $N$-parameter ($N\ge 3$) tight-binding models for several cuprate families\cite{1band}, and shown that they can be applied to a many-body formalism to  predict material-specific properties, as long as the pseudogap is predominantly a short-range antiferromagnet\cite{AIP}.  More recently, we have found that by taking a more approximate 3-parameter parametrization for the cuprates, we can recover the same phase diagrams by tuning only a single parameter t’ for each cuprate family\cite{MBMB}.  Using this parametrization, we have interpolated between cuprates to uncover a Mott-Slater transition, which we believe plays a key role in underdstanding the doping dependence of cuprates.  The key result of Ref.~\onlinecite{gapsend} is to reveal the experimental signatures of this transition.  Here we use these parameter sets to successfully describe the experimental doping dependence of several properties characterizing pseudogap collapse.  

We refer to the 3-parameter tight-binding model as the reference family for cuprates, and discuss here how it is determined.  The parameters are the nearest ($t$), next-nearest ($t'$), and next-next-nearest ($t''$) hopping parameters on a square lattice of copper atoms.  The cuprate family is specified by $t''=-t'/2$\cite{PavOK,AIP}.  For the La and Bi cuprates, the $t'$ parameter values were determined to ensure that the susceptibility of the reference family matched that of the more accurate $N$-parameter models.\cite{MBMB}  To generalize to other cuprates, we had hoped to adopt the Pavarini-Andersen parameters\cite{PavOK}, but found that these values were too large, particularly for cuprates with larger $t'$-values.\cite{hoVHS2a}  Here we clarify the issue with the Pavarini-Andersen model, and show how a scaled version of their results can be used.

Pavarini {\it et al.}\cite{PavOK} introduced a parameter $r$, and showed that to lowest order 
 \begin{equation}
t'/t=-br,
\label{eq:A1} 
\end{equation}
with $b=1$, where
 \begin{equation}
r=\frac{1}{2}c_n,
\label{eq:A2} 
\end{equation}
$c_n=cos(k_{xnV}a)$, and $k_{xnV}$ is the $x$-component of the $k$-vector from the origin $(\Gamma)$ to the Fermi surface in the nodal ($\Gamma\rightarrow (\pi,\pi)$) direction at $x_{VHS}$.\cite{thankOKA}  For arbitrary values of $t''$, $r$ can be found from
 \begin{equation}
tc_n+(t'+2t'')c_c^2=-(t'-2t''),
\label{eq:A2} 
\end{equation}
which reduces to Eq.~\ref{eq:A1} with $b=1$ when $t''=-t'/2.$  Thus, if the $t-t'-t''$ model provided a perfect fit to the true dispersion, the value of $t'$ could be determined by the measurement of a single $k$-value.  In reality, the fit is not perfect, so one needs to choose $t'$ to give the best fit to the full dispersion.  This leads to an offset between $t'$ and the value determined by $r$.  We propose that that offset can be accounted for by optimizing $b$ in Eq.~\ref{eq:A1}.

\begin{figure}
\leavevmode
\rotatebox{0}{\scalebox{0.50}{\includegraphics{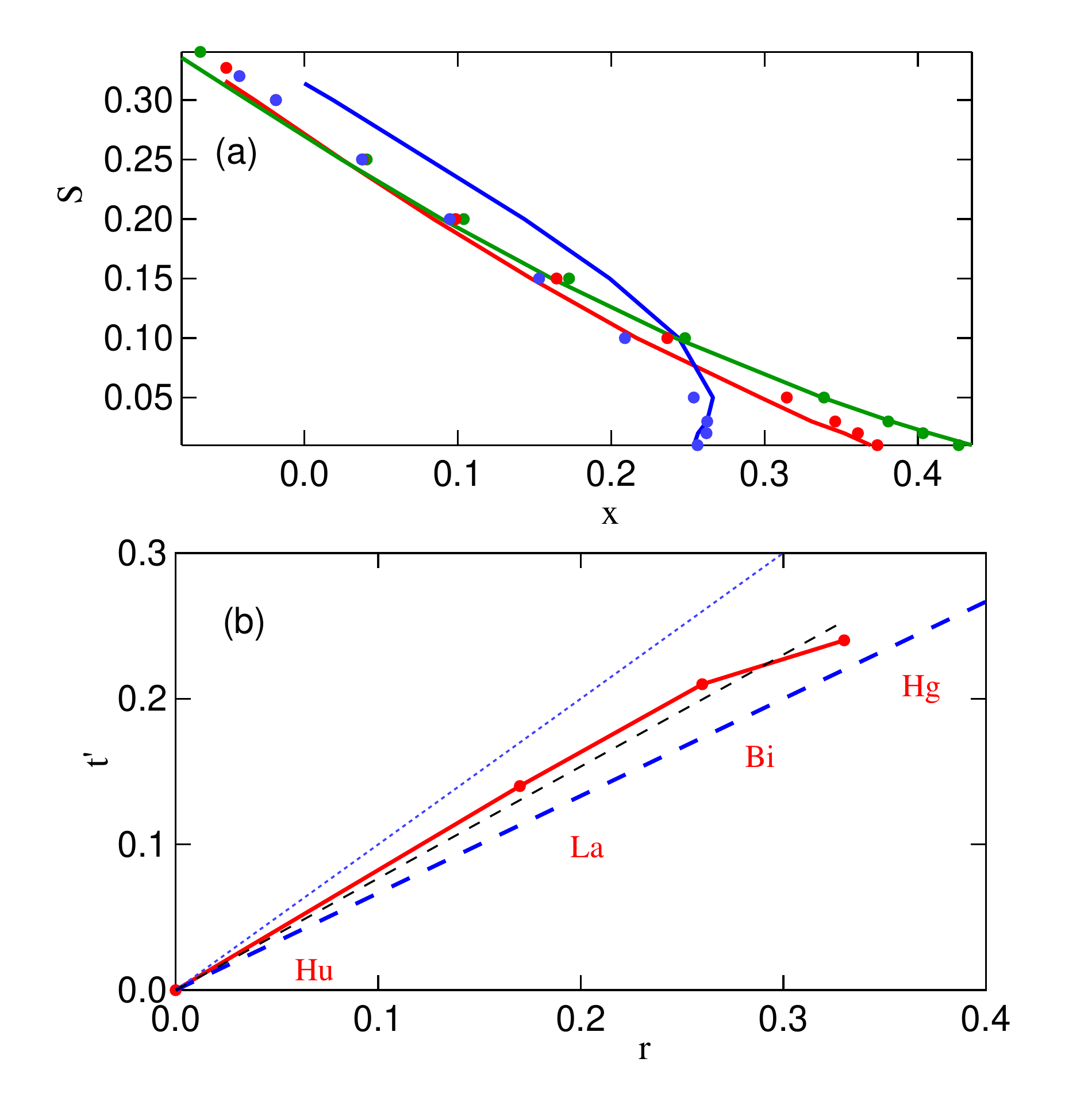}}}
\vskip0.5cm
\caption{
(a) Comparison of more accurate $N$-parameter tight binding models (solid lines) to the $3$-parameter reference family (dashed lines) in ability to predict $S(x)$ phase diagram of cuprates.  Solid lines represent fits to renormalized ($Z=0.5$) DFT dispersion of LSCO (blue)\cite{AIP} and Hg cuprates (green)\cite{Tanmoy} and to experimental dispersion of Bi2201\cite{JennySci}, and filled circles represent cuprate reference family ($t''=-t'/2$) with $t'/t$ = -0.14 (blue),  -0.21 (red), and -0.24 (green).  (b) Comparison of best-fit $t'$ from (a) with parameter $r$ from Ref.~\onlinecite{PavOK} (red circles), for the three materials above (La, Hg, and Bi respectively) and for the Hubbard model (Hu).  Blue dashed (dotted) lines represent the relations $t'/t=-2r/3$ ($t'/t=-r$).
}
\label{fig:A0}
\end{figure}

Here, we perform additional calculations to test this hypothesis.  We take three $N$-parameter ($N>3$) tight binding fits from earlier calculations\cite{AIP,Tanmoy,JennySci} and recalculate the magnetization vs doping for them, solid lines in Fig.~\ref{fig:A0}(a).  Then, we find the $t'$ value of the reference family which best reproduces this data, dashed lines.  The goodness of the fits supports our idea of using the reference family. In Fig.\ref{fig:A0}(b) we plot these values against the Pavarini-Andersen $r$-values for the same materials, including the Hubbard model, where both must have $t'=r=0$. The best fit to the data (black short-dashed line) has $b=0.767\pm0.025$.  In this paper we will use the $r$ values of Ref.~\onlinecite{PavOK} and set $b=0.767$.  These results are summarized in Table~SI, where the top row ($b=1$) lists the Pavarini-Andersen values ($b=1$ in Eq.~\ref{eq:A1}), and the next row ($b=0.767)$ gives the best fit to the $N$-parameter fits (third row).  Instead of fitting to earlier models of the dispersions, we could directly fit to the experimental data in Fig.~1, by finding the values of $t'$ that give the best agreement with the experimental values of $x^*$ (fifth row of Table I).  The choice $b=2/3$ (fourth row) give the best fit.  References to the experimental data are given in the Fig.~1 figure caption.

\begin{center}
\begin{table}
\begin{tabular}{ |c|c|c|c| } 
 \hline
  & La & Bi & Hg [Tl] \\ 
 \hline
b=1 & -0.17 & -0.26 & -0.33, -0.32[Tl] \\  
 \hline
b=0.767 & -0.13 & -0.20 & -0.253 \\ 
N-par. & -0.14 & -0.21 & -0.24 \\  
 \hline
b=2/3 & -0.113 & -0.173 & -0.213[Tl] \\ 
Expt. & -0.08, -0.12 & -0.18, -0.21 & -0.18[Tl] \\
 \hline
\end{tabular}
\caption{Alternative choices of $b$ in Eq.~\ref{eq:A1}.}
\label{table:S1}
\end{table}
\end{center}


{\it A technical note.}  In fitting a tight-binding model to the Cu-$d_{x^2-y^2}$ antibonding band, one should note that away from the Fermi surface this band overlaps and hybridizes with several other bands.  Thus, one should fit the model to the spectral weight of the Cu-$d_{x^2-y^2}$ orbital character, which is spread over several bands.  Fitting directly to experimental data is more challenging, since one cannot in general extract the orbital character of the bands, and the bands are broadened, which can only be included in theory by a self-energy correction.

\subsection{Details of the calculations of Fig. 1}

We calculate the AFM doping dependence via a self-consistent (in density and magnetic moment) Hartree-Fock calculation, informed by many-body perturbation theory and {\it ab initio} calculations.  Thus, analysis of the QPGW self-energy\cite{AIP}  finds that (1) while the Hubbard $U\sim$ 2eV for undoped cuprates, with doping it is screened by long-range Coulomb fluctuations, falling very rapidly or discontinuously to $\sim$1~eV; (2) for finite doping the coherent dispersion is renormalized by a factor $Z\sim$0.5, which also renormalizes $U$.  (3) The renormalization turns on below an incoherent-to-coherent crossover energy/temperature scale associated with a peak in the imaginary self-energy.  By confining our results to the low temperature coherent regime, we can work with fixed $U=3t$, $Z=0.5$ (solid lines in Fig.~1), with a generic bare $t$=0.42eV.

\begin{figure}
\leavevmode
\rotatebox{0}{\scalebox{0.40}{\includegraphics{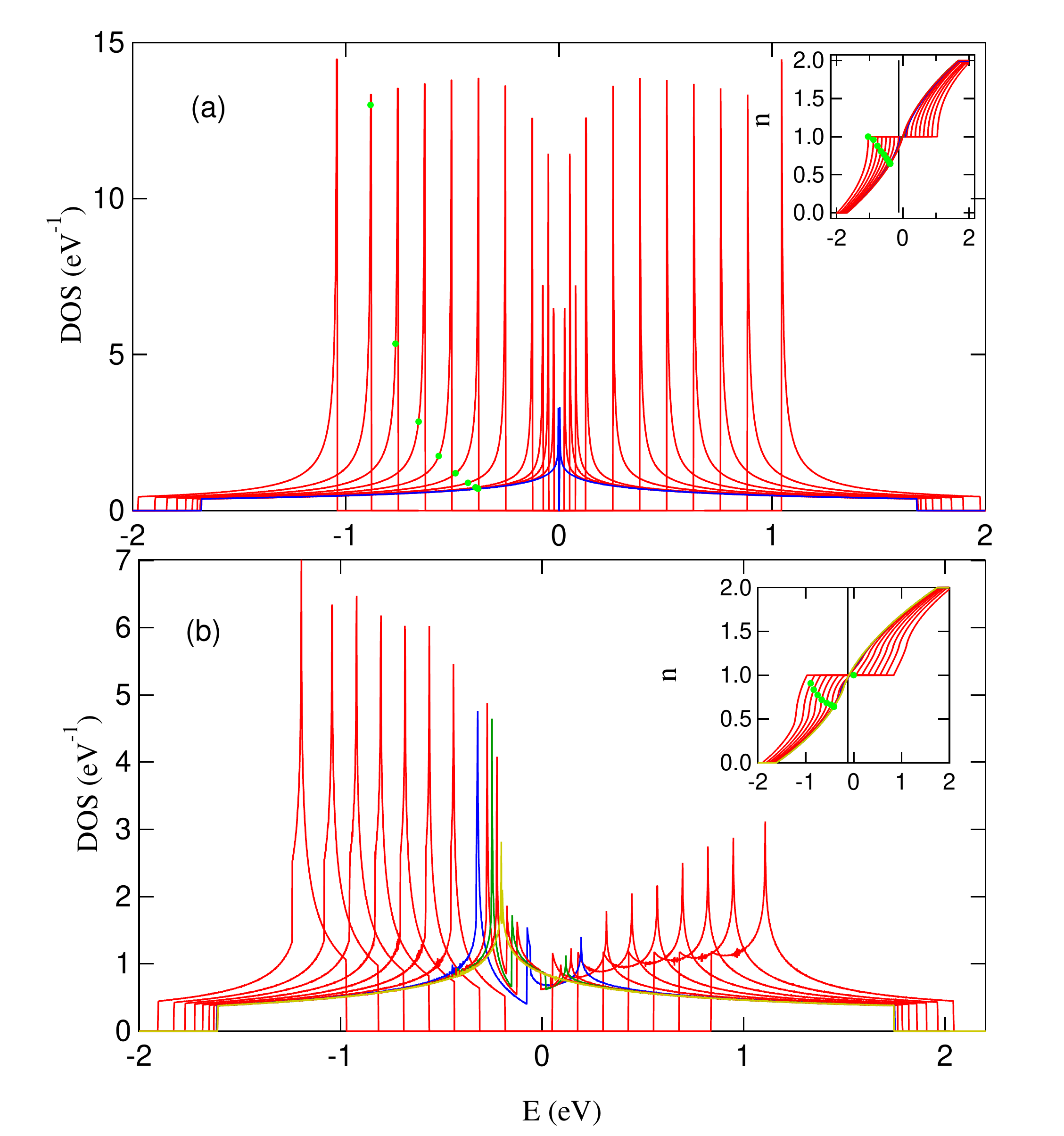}}}
\vskip0.5cm
\caption{
Merging of VHSs of upper and lower magnetic bands at AFM collapse. Blue curves indicate initial gap closing.  Insets show corresponding electron density $n$, with green dots indicating Fermi level.
}
\label{fig:80a}
\end{figure} 

In Fig.~1(a), we define a dimensionless magnetization $S$ by $\Delta=US$, $S=<n_{\uparrow}-n_{\downarrow}>/2$, and the average $<...>$ is over occupied $k$-states.  The efficacy of our self energy correction is clear: at low doping all curves converge to $S=0.3$, giving an average magnetic moment $\mu=2S\mu_B=0.6\mu_B$, where $\mu_B$ is the Bohr magneton, in good agreement with experiment\cite{Tranq2,Kanun}, while the gap parameter is $2\Delta = 0.76$eV, close to the experimental 1eV.\cite{Ando,YUBOI}  In contrast, the bare $U=6t$ and $Z=1$ (long dashed lines) lead to too large  values for $S$.

\begin{figure}
\leavevmode
\rotatebox{0}{\scalebox{0.46}{\includegraphics{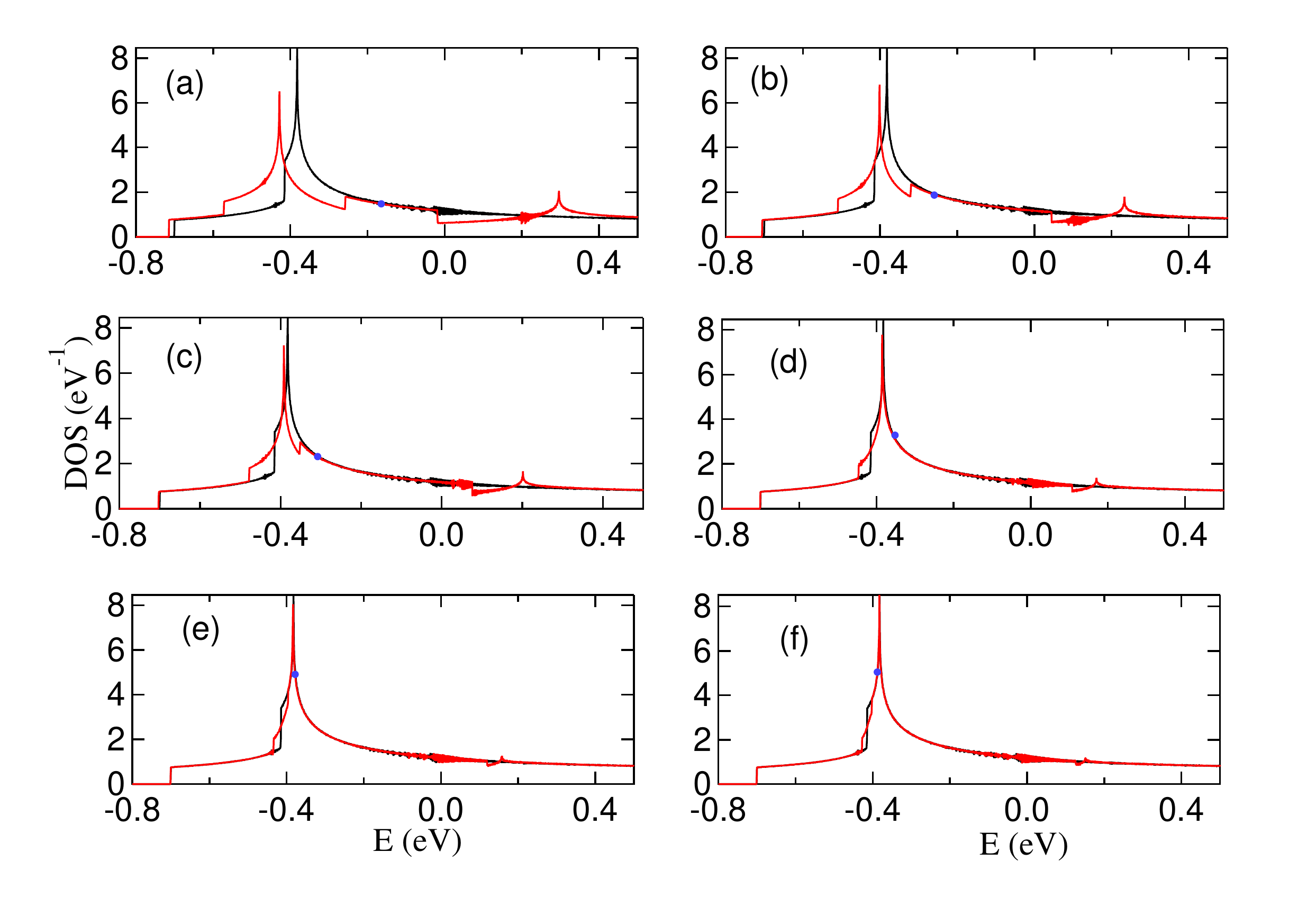}}}
\rotatebox{0}{\scalebox{0.46}{\includegraphics{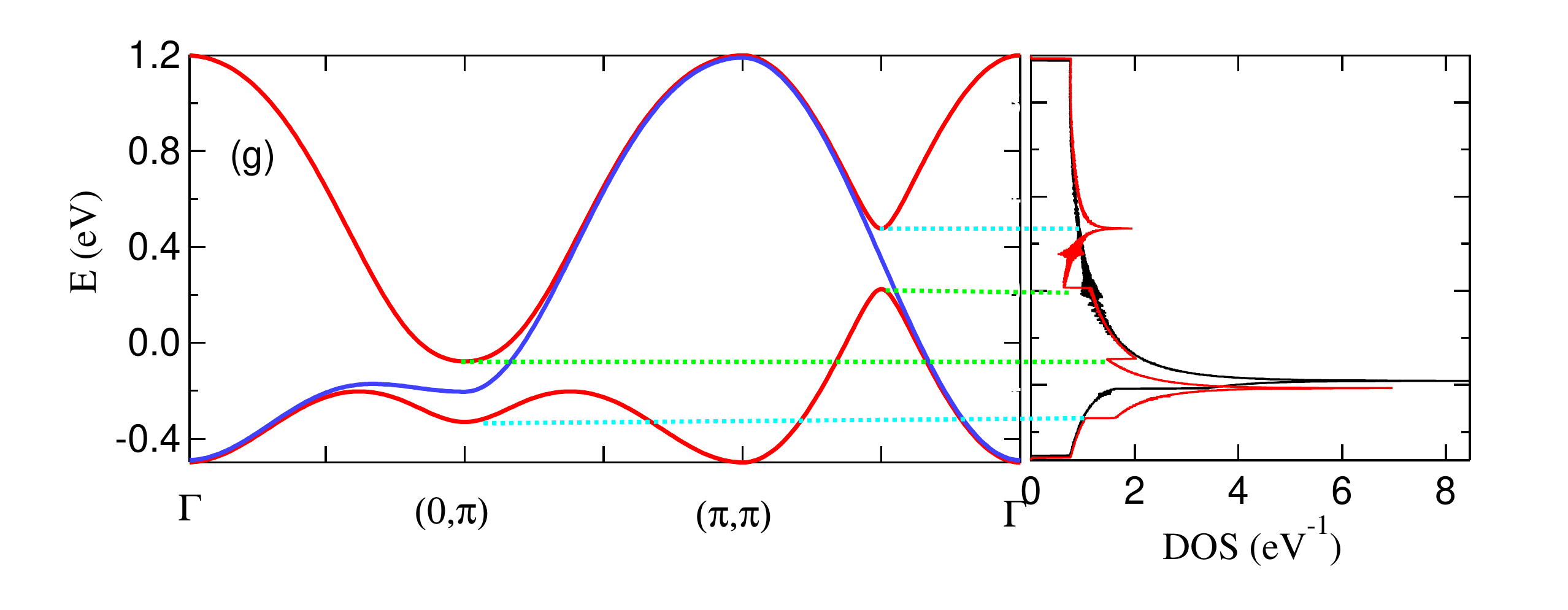}}}\vskip0.5cm
\caption{
{\bf Hidden phase transition}.  (a-f) DOS vs energy for increasing doping in AFM phase (red curves) vs nonmagnetic phase (black curves), with $S$ = 0.25 (a), 0.15 (b), 0.1 (c), 0.05 (d), 0.03 (e), and 0.02 (f).  Blue diamond indicates Fermi energy of AFM phase.    (g)  Dispersion vs DOS for data of frame (a).  Blue and green lines relate features in the DOS to underlying features in the dispersion.
}
\label{fig:80}
\end{figure} 

As the AFM gap collapses, a key role is played by the VHSs of the upper and lower magnetic bands (U/LMBs), Fig.~\ref{fig:80a}, and how they evolve with doping.  For the original Hubbard model with $t'=0$, Fig.~\ref{fig:80a}(a), the process is simple.  The VHS of the LMB falls at the top of the band and the VHS of the UMB falls at band bottom.  With doping the DOSs of the two bands approach one another virtually without change of shape, but with gradually reducing amplitude, until they merge into a single peak as the AFM magnetization $S\rightarrow 0$, Fig.~\ref{fig:80a}(a).  The need for a self-consistent calculation can be understood from this figure.  The gap is stabilized because the electronic kinetic energy is lowered in the gapped phase, and most of that energy lowering is associated with electrons near the VHS of the LMB.  Since the magnetic VHS is at the top of the LMB, hole-doping the sample moves the fermi level below the VHS, and the large electronic stabilization energy is quickly lost, causing the gap to collapse. 

The case for finite $t'$ is much more interesting, Fig.~\ref{fig:80a}(b).  With increasing doping, spectral weight in the UMB shifts from the saddle point VHS to the leading edge, causing it to peak at the bottom of the UHB, while the VHS of the LMB grows stronger as it moves toward the top of the band.  At the same time, the two bands approach one another, until as $S\rightarrow 0$, the leading edge of the UMB merges with the saddle-point VHS of the LMB to form a single peak.  In contrast, the saddle-point VHS of the UMB shrinks and merges with the leading edge of the LMB to form a featureless background.  As seen in Fig.~1, the LMB VHS is in general distinct from the VHS of the nonmagnetic (NM) band, with the gap closing at a doping higher than $x_{VHS}^{NM}$.

The effect of this VHS collapse on the AFM order depends sensitively on the magnitude of $U$, Fig.~1.  In Figure~1, the black curves (for $t'=-0.08t$) compare the doping evolution for several parameter values: unrenormalized $U=6t$, $Z$=1 (long-dashed curve) vs $Z=0.5$ and either $U=3t$ (solid line) or $U=1.5t$ (short-dashed line).  For the first case, the NM VHS (thin black line in Fig.~1(b)) is far from the Fermi level when $S\rightarrow 0$, so the NM VHS has little effect on the transition.  However, for the smaller $U$, the AFM VHS first approaches the NM VHS, when $U=3t$, and crosses it near $U=1.5t$.  Note that whereas the NM VHS has a weak logarithmic divergence, 
the AFM VHS is close to a high-order VHS with strong power-law divergence.  Note further that the gap collapse is first-order, as seen by the nonmonotonic $S(x)$ curve.  

The remaining curves in Fig.~1 illustrate the strong evolution of the magnetic moment decay and the DOS with $t'$.  We note that the first-order transition changes to second-order around $t'=-0.2t$.  The $t'= -0.32$ and $-0.33t$ data are particularly interesting.  There is a strong magnetic moment which decays to zero, but with {\it no sign of any transition in the DOS, which is identical to the nonmagnetic case.}  This is a remarkable example of a hidden phase transition associated with a VHS, as illustrated in Fig.~\ref{fig:80}.  The energy lowering arises from pushing the VHS peaks away from the Fermi level, but at low energies the two magnetic bands overlap (green dotted lines in Fig.~\ref{fig:80}(g)), creating a DOS that is virtually indistinguishable from the nonmagnetic DOS.

In the above discussion, we showed that AFM order persists to the largest $|t'/t|$ studied.  However, for such large $|t'|$, there can be a competition between inter- and intra-VHS scattering, with the latter leading to a more conventional form of CDW.  The latter is predicted to dominate for $t'/t \le -0.23$,\cite{hoVHS1} but so far there is little evidence for it. 

\subsection{AFM VHS line shape}
\begin{figure}
\leavevmode
\rotatebox{0}{\scalebox{0.40}{\includegraphics{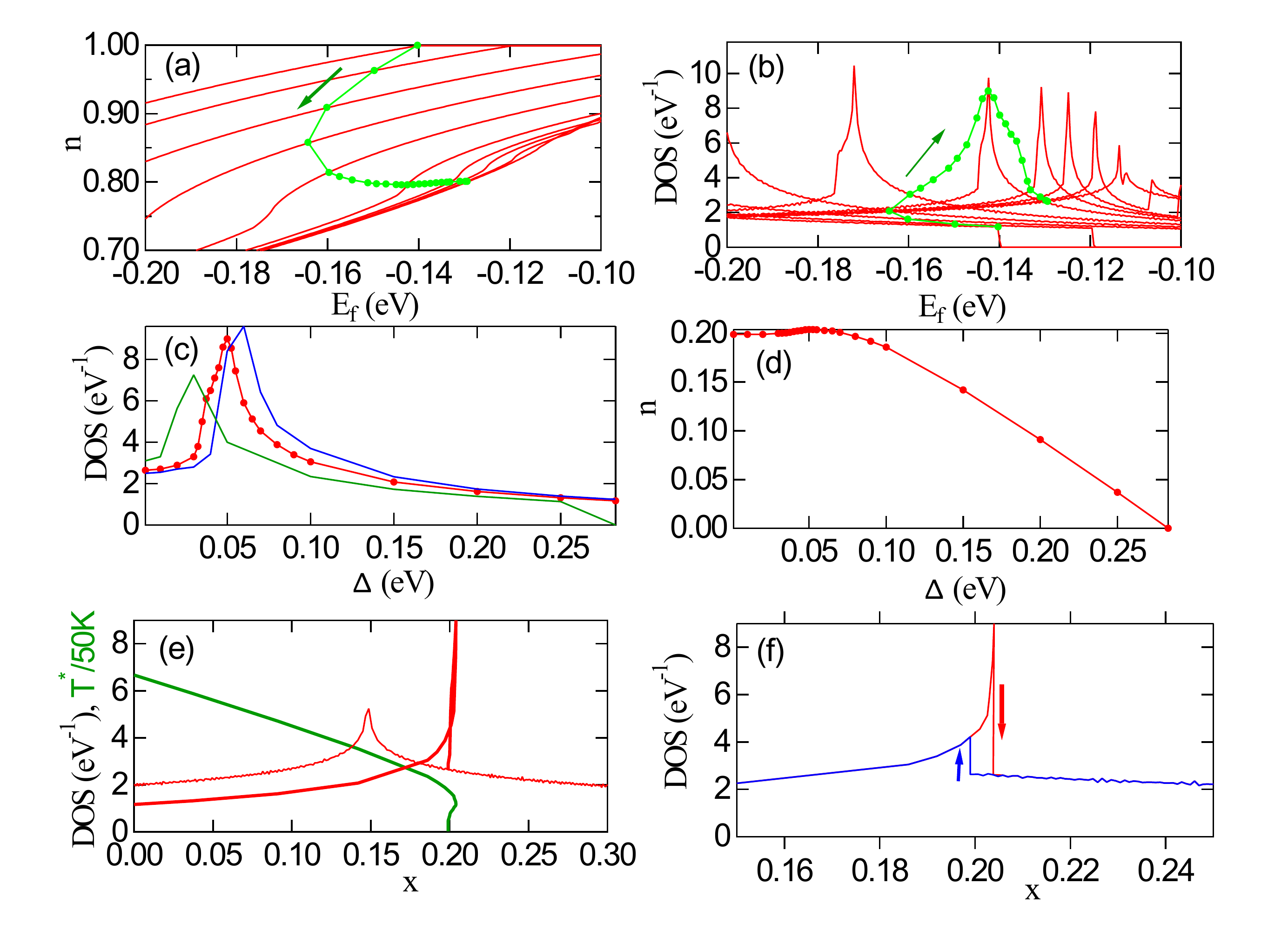}}}
\vskip0.5cm
\caption{
{\bf AFM DOS near $x_{pg}$}, for $t'=-0.09t$.  (a,b)  Band filling $n$ (a) and DOS (b) evolution with gap parameter $\Delta$, where green dots represent self-consistent $E_f$ for different $\Delta$, and the green arrows indicate the direction of decreasing $\Delta$ (equivalently, increasing hole doping).  (c,d) Resulting DOS (c) and hole doping $x$ (d) vs $\Delta$.  In (c), lower-resolution DOSs are shown for $t'/t$ = -0'08 (blue) and -0.12 (green).  (e)  DOS (red) and $T^*$ (calculated by assuming $2\Delta/(k_BT^*)=X$ = 24.8 to  approximate experiment) vs $x$.  The nonmonotonic evolution of $T^*$ and $\Delta$ indicates a first order transition.  (f)  Corrected DOS in case of maximum hysteresis, with arrows indicating sweep directions.}
\label{fig:9a}
\end{figure}
In Fig.~\ref{fig:9a}, we study the first-order AFM transition in more detail, using $t'/t=-0.09$ as an example.  For each gap parameter $\Delta$, the $T=0$ dispersion is calculated and used to find the fermi energy at which the self-consistency equation for $\Delta$ is satisfied, from which the band filling $n$ and the DOS are calculated, frames (a,b).  Each green dot indicates a separate calculation at a different $\Delta$.  Note that as the AFM VHS approaches the Fermi level, it pushes the value of $n$ backwards, like a wave pushing a floating log back towards the shore.  In frames (c,d), we see that the DOS and doping $x=1-n$ are smooth functions of $\Delta$.  However, $\Delta$ and the resulting $T^*$, green line in frame (e), are not monotonic functions of $x$, and the regime where $\Delta(x)$ has three values indicates a first-order instability.  Frame (f) illustrates the maximum possible hysteresis one could have, showing that part of the DOS peak is cut off.  Indeed, since hysteresis is not observed, the resulting first-order step should be calculated by a Maxwell construction, which would lead to a step roughly half way between the two hysteretic branches, in which case most of the DOS peak would be lost.  

Thus, while the presence of an enhanced VHS at the AFM transition and the sharp step in $T^*$ are both individually in good agreement with experiment, something goes wrong when we try to put them together: the step cuts off much of the DOS peak.  Is there a way out of this problem?  The strong swings in $E_f$, frames (a,b) are characteristic of an isolated system, lacking a particle reservoir.  But in a system with NPS, such as the gap maps, the other patches can serve as a reservoir.  Indeed, since each patch corresponds to a particular gap, each patch near $x_{pg}$ will have a particular DOS, frame (c).  The quantum confinement in patches will also freeze out the backbending of the hole doping, frame (e).  In this case, STM experiments can resolve patches with individual gap values, while macroscopic experiments should find an average gap that evolves smoothly with doping, leading to a DOS peak similar to the peaks in frame (b) (green dots), but broadened out by the distribution of oxygen environments.  
As discussed in Section III, this broadened DOS captures the key features of the experimental heat capacity.\cite{Michon}

\subsection{c-axis dispersion}
Recently, additional evidence was adduced that the observed heat capacity peak at pseudogap collapse could not be associated with the normal state VHS in LSCO.  ARPES finds a c-axis contribution to the electronic dispersion, of peculiar form associated with the body-centered tetragonal (BCT) form of Cu stacking in adjacent planes.\cite{LSCO3D}  This c-axis dispersion cuts off the divergence of the VHS peak altogether, leaving behind a flat topped peak.\cite{LSCO3D,ZXLifshitz}

Here we note that complications of c-axis dispersion should be much weaker in the AFM phase.  First, in the pure AFM phase, the BCT stacking frustrates interlayer magnetic coupling, as a given Cu on one layer has equal numbers of nearest neighbor up and down spins on the next layer.  Secondly, the stripe phase in LSCO has an unusual period quadrupling along the c-axis.  This is generally interpreted in terms of coupling to (usually fluctuating) regions of low-temperature tetragonal order, which causes the stripe order to run alternately along the a- or b-axes in adjacent layers.\cite{Tranq}  The remaining period doubling is due to the parallel charge stripes on every other layer shifting laterally to remain as far apart as possible to minimize Coulomb repulsion.  Both of these effects are readily seen to reduce interlayer hopping, particularly since the doped holes are mainly confined to the charged stripes.

Finally, we note that in the related compound La$_{2-x}$Ba$_x$CuO$_4$, the stripes have long-range order, which quenches superconductivity, particularly near $x=0.125$.  In the vicinity of the stripe order, strong {\it two-dimensional} superconducting fluctuations have recently been discovered.\cite{Tranq2D}

\section{Low-field transport and Hall effect}

\begin{figure}
\rotatebox{0}{\scalebox{0.66}{\includegraphics{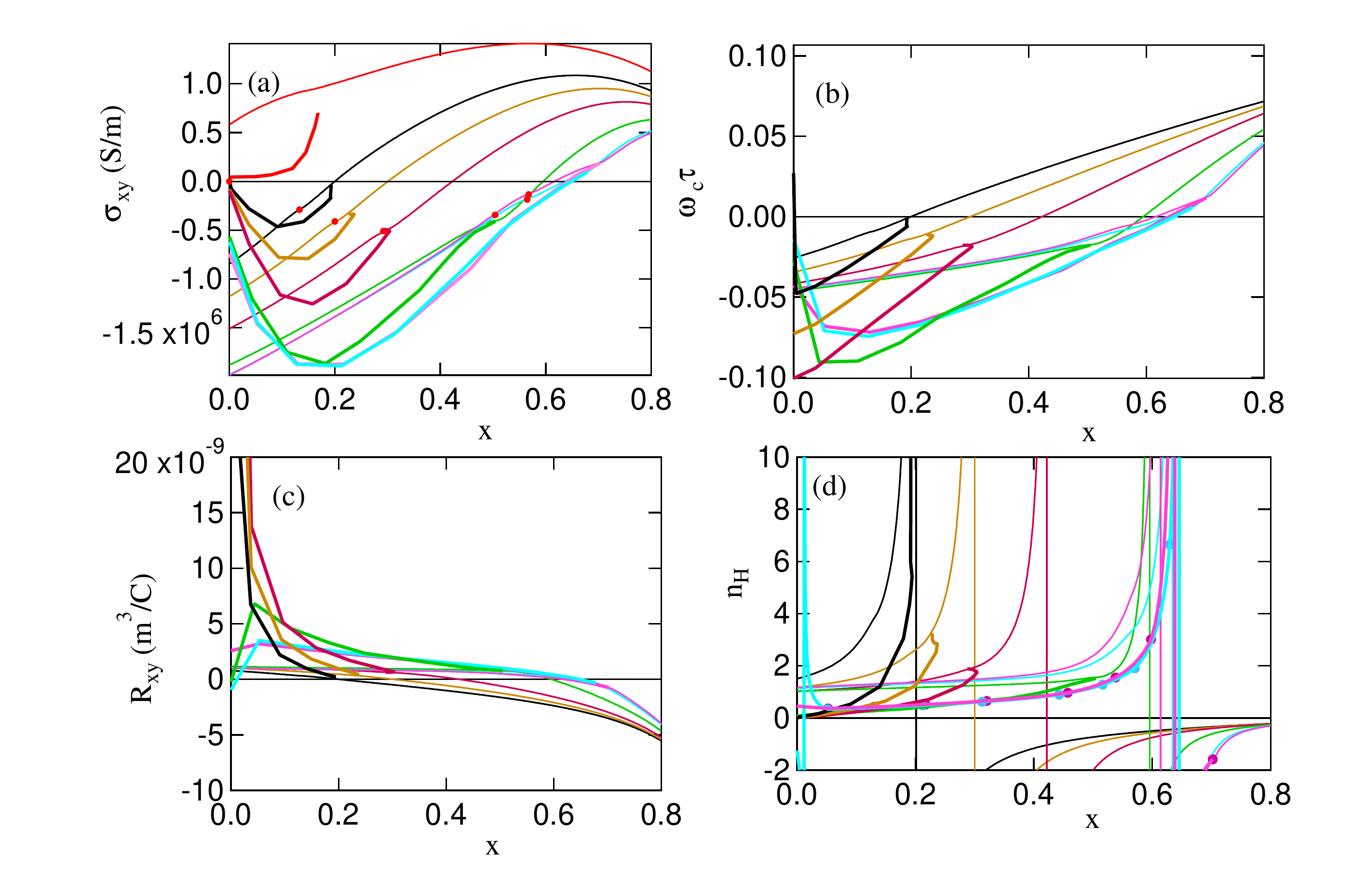}}}
\vskip0.5cm
\caption{
Low-field transport for the $t-t'-t''$ model of cuprates, for several values of $t'$, showing (a) $\sigma_{xy}$, (b) $\omega_c\tau$, (c) $R_{xy}$, and (d) $n_{H}$.  Color scheme of curves is consistent with Fig.~1.  
}
\label{fig:77}
\end{figure} 

\begin{figure}
\rotatebox{0}{\scalebox{0.5}{\includegraphics{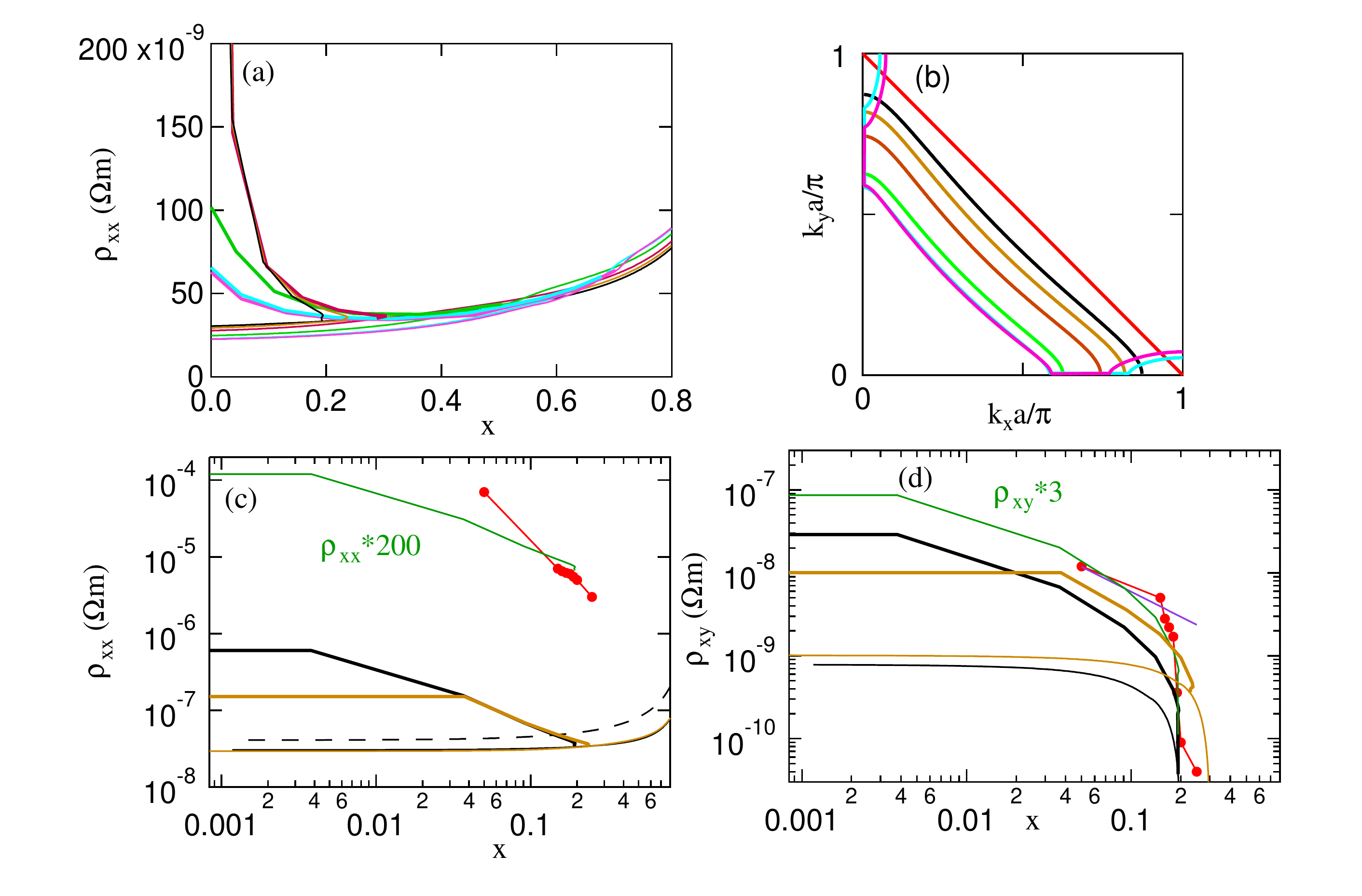}}}
\vskip0.5cm
\caption{
Low-field Hall effect compared to experiment, showing (a) $\rho_{xx}$ and (b) Fermi surfaces at the Hall zeroes for the same parameter sets as in Fig.~\ref{fig:77}.  Frames (c-d) continue the comparison of theory and experiment for LSCO\cite{OngHall} (red dots) in Fig.~4(b). 
}
\label{fig:78}
\end{figure}

In the main text Section V, we discuss low-field transport properties, focusing on the conductivity $\sigma_{xx}$ and Hall number $n_H$.  Here we briefly discuss additional transport properties.  The main result of this calculation is that for each $t'$ there is a doping $x_{H0}$ at which the Hall conductivity $\sigma_{xy}$ = 0, Fig.~\ref{fig:77}(a).  Consequently, the Hall angle $\omega_c\tau = \sigma_{xy}/\sigma_{xx}$, Fig.~\ref{fig:77}(b) and resistivity $\rho_{xy}= -\sigma_{xy}/(\sigma_{xx})^2=R_HH$, Fig.~\ref{fig:77}(c), with $R_H=1/(n_He)$, also pass through zero at the same doping, so that $n_{H}\rightarrow\infty$, Fig.~\ref{fig:77}(d).  
Since this crossover involves a change from hole-like to electron-like conductivity, one might infer that it happens at the VHS, $x_{H0}=x_{VHS}$.  One would be wrong.  The low-field Hall conductivity is insensitive to the global nature of the Fermi surface -- electron-like or hole-like -- since $\omega_c\tau < 1$.  Instead, the Hall conductivity is sensitive to the local curvature, and the Hall zero arises when the Fermi surface has equal amounts of positive and negative curvature.  This is illustrated in Fig.~\ref{fig:78}(b), which displays the Fermi surfaces at each Hall zero.  The red dots in Fig.~\ref{fig:77}(a) represent $x_{VHS}^{NM}$, showing that $x_{H0}=x_{VHS}^{NM}$ only in the Hubbard limit $t'=0$.

Figures~\ref{fig:78} and~4(b,c) show a key result, a comparison of our predicted resistivity and Hall density with the measured\cite{OngHall} Hall resistivity in LSCO, compared to theory for $t'/t=-0.08$ or $t'/t = -0.12$.  Figures~\ref{fig:78}(a) and ~\ref{fig:77}(d) show the evolution of $\rho_{xx}$ and $n_H$ with $t'$, while Figs.~\ref{fig:78}(c,d) and~4(b) are specific to LSCO.   Our calculated resistivity is in good agreement with  $\rho_{xx}$ calculated for the commonly used model $\sigma_{xx} =ne^2\tau /m$ (black dashed line in frame (c)), with $m=4m_0$, and $m_0$ is the free electron mass.  By choosing a constant $\tau=\hbar/\gamma_0$, with $\gamma_0$=1~meV, we can extract the experimental scattering rate $\gamma$ in meV by taking the ratio of the experimental resistivity to the calculated value.  The average $\gamma=200$~meV found in frame (c) suggests strong scattering associated with stripe physics.  In contrast, $\rho_{xy}$ should be independent of a constant $\tau$, and in Figs.~\ref{fig:78}(d), ~4(b,c) we find much better agreement with experiment.  

In Figs.~4(b) and~\ref{fig:78}(d) we show two different theoretical calculations, one for $t'=-0.08t$, which captures the steep drop in $R_H$ (Fig.~\ref{fig:78}(d)), but underestimates the low doping values by a factor of 3, and the other for $t'=-0.12t$, better fitting the low-doping regime and the expected zero-crossing.  By substituting a rare earth atom for La, or Ba for Sr, a large variety of LSCO-like compounds can be formed, which have different values for $x_{pg}$,\cite{Michon} presumably due to different values of $t'/t$, as in Fig.~\ref{fig:78}. 

The steep rise of the experimental $n_{H}$ near $x=0.2$, followed by a slower variation without clear change of sign at higher doping, is suggestive of significant disorder effects -- possibly associated with strong VHS scattering giving rise to a flat-band.\cite{Nazar2,BulutSW}  In these early experiments sample quality could be an issue, so further experiments at higher doping would be helpful in pinning down the theoretical values.  In passing, we observe that (a), the shift of $n_H$ to the $1+x$ line (violet dashed lines in Figs.~2(b,c)) is not clearly observed in the data, and (b) the NM Hall data show deviations from $1+x$ unless $x_{VHS}$ is quite large, e.g., for the calculated Bi2201 curve (thin green line).   
